\documentclass[12pt]{article}

\usepackage{a4wide}
\usepackage[dvips]{graphicx}
\usepackage{amsmath,amsfonts}

\newcommand{\md}{{\rm d}}
\newcommand{\sig}{\textrm{sgn}}
\newcommand{\n}{\bar{\nu}}

\newtheorem{theo}{Theorem}
\newtheorem{defi}{Definition}
\newtheorem{lemma}{Lemma}

\newcommand{\proofend}{\raisebox{1.3mm}{\fbox{\begin{minipage}[b][0cm][b]{0cm}
\end{minipage}}}}
\newenvironment{proof}{\noindent{\it Proof:} }{\mbox{}\hfill
\proofend\\\mbox{}}

\begin{document}

{\renewcommand{\thefootnote}{\fnsymbol{footnote}}
\hfill  AEI--2005--170, NI05064\\
\medskip
\hfill gr--qc/0511058\\
\medskip
\begin{center}
{\LARGE Perturbative Degrees of Freedom in\\[2mm] Loop Quantum Gravity: Anisotropies}\\
\vspace{1.5em}
Martin Bojowald\footnote{e-mail address: {\tt mabo@aei.mpg.de}},
H\'ector H.\
Hern\'andez H.\footnote{e-mail address: {\tt hehe@aei.mpg.de}}
\\
\vspace{0.5em}
Max-Planck-Institut f\"ur Gravitationsphysik, Albert-Einstein-Institut,\\
Am M\"uhlenberg 1, D-14476 Potsdam, Germany\\
\vspace{0.7em}
Hugo A.\ Morales T\'ecotl\footnote{e-mail address: {\tt hugo@xanum.uam.mx},
Associated member of ICTP, Trieste Italy.} \\
\vspace{0.5em}
Departamento de F\'{\i}sica, Universidad Aut\'onoma
Metropolitana Iztapalapa\\ A.P.~55--534, M\'exico D.F.
09340, M\'exico
\vspace{1.5em}
\end{center}
}

\setcounter{footnote}{0}

\begin{abstract} 
The relation between an isotropic and an anisotropic model in loop
quantum cosmology is discussed in detail, comparing the strict
symmetry reduction with a perturbative implementation of
symmetry. While the latter cannot be done in a canonical manner, it
allows to consider the dynamics including the role of small
non-symmetric degrees of freedom for the symmetric evolution. This
serves as a model for the general situation of perturbative degrees of
freedom in a background independent quantization such as loop quantum
gravity, and for the more complicated addition of perturbative
inhomogeneities. While being crucial for cosmological phenomenology,
it is shown that perturbative non-symmetric degrees of freedom do not
allow definitive conclusions for the singularity issue and in such a
situation could even lead to wrong claims.
\end{abstract}

\section{Introduction}

By far the most common and extended studies in cosmological models are
those admitting only finitely many or even one gravitational degree of
freedom, the scale factor or radius of the universe $a(t)$ and maybe
anisotropy parameters. This is justified in a homogeneous or isotropic
model, symmetries present in our universe at large scales. This is
also mainly implemented in models of quantum gravity, such as in
Wheeler--DeWitt quantum cosmology \cite{DeWitt,QCReview} and loop
quantum cosmology \cite{LoopCosRev,WS:MB,LivRev}. That degree of symmetry
does not only allow insights into phenomenological aspects of the
early universe but, in anisotropic models, also displays the most
crucial features of classical singularities.

More realistic models can then be obtained by breaking some of the
symmetries imposed as before. After isotropy the first step is
anisotropy, which can often be implemented exactly, and then
inhomogeneity as the crucial one. In the first case (anisotropy) there
is still a finite number of degrees of freedom and usually they do
not pose much new technical difficulties. Conceptually, however, they
are essential in particular for a general understanding of the issue
of singularities (see, e.g., \cite{DegFull} for a recent discussion in
the context of quantum cosmology). It is then of interest to discuss
anisotropic models at different levels and to see what the
implications for singularities are. Complete quantizations have been
done in \cite{HomCosmo,Spin,BHInt} with the result that their loop
quantizations are free of singularities as in isotropic cases
\cite{IsoCosmo,Sing,Bohr}.

Alternatively, one can try to implement anisotropies as perturbations
to a quantized isotropic model and see how correction terms change the
dynamical behavior.  In this work we develop such a point of view both
to shed light on the singularity issue and as a model for perturbative
inhomogeneities. Following \cite{HomCosmo} where anisotropic models
have been quantized in the loop framework, we introduce suitable
anisotropy parameters in that model and implement conditions for them
to be small. (As a first step we take only one anisotropy parameter of
two possible ones which means that there is still a single rotational
symmetry axis, a so-called locally rotationally symmetric (LRS)
situation.) This will be discussed in detail at the kinematical and
dynamical levels and results in an evolution equation describing an
anisotropic model perturbatively around an isotropic one. The
formulation can also be used for comparing background independent
theories with theories formulated on a background, because the
isotropic model can be seen as a background for the perturbative
anisotropy.

In section 2 we describe the classical models, to be quantized in
section 3 both exactly and in a perturbative manner. This illustrates
how symmetric models can be related to a less symmetric theory at the
quantum level.  In section 4 we discuss the evolution equation in
terms of the perturbation and its new features compared to the
isotropic model. In section 5 we look at the singularity issue from
the perturbative point of view and show that conclusions drawn in such
a situation must be used much more carefully than in a
non-perturbative quantization.

\section{Classical models}

For simplicity, we consider the two most simple models, which are the
flat isotropic model and the Bianchi I model with one additional
rotational symmetry (LRS). Since loop quantizations are based on
Ashtekar variables \cite{AshVar,AshVarReell} given by a connection
$A_a^i$ and a densitized triad $E^a_i$ we first introduce these
variables. In models, they take special forms corresponding to the
fact that they are required to be preserved by a symmetry
transformation up to a gauge transformation.

An isotropic connection and densitized triad for a flat model can
always be written as
\begin{eqnarray}
A_a{}^i \md x^a &=& \bar{c} \md x^i\\
E_i{}^a \frac{\partial}{\partial x^a}&= & \bar{p}
\frac{\partial}{\partial x^i}\,.
\end{eqnarray}
in Cartesian coordinates, where the only two remaining components are
conjugate to each other,
\begin{equation}
 \{\bar{c},\bar{p}\} = \frac{8\pi}{3}\gamma G
\end{equation}
with the Barbero--Immirzi parameter $\gamma$
\cite{AshVarReell,Immirzi} and the gravitational constant $G$. The
variable $\bar{p}$ contains information about spatial geometry such as
the volume
\begin{equation}
 V_{\rm iso}= |\bar{p}|^{3/2}
\end{equation}
while $\bar{c}$ is proportional to extrinsic curvature. Both variables are
related dynamically as dictated by the Hamiltonian constraint
\begin{equation}
 H_{\rm iso} = -\frac{3}{8\pi G\gamma^2}\bar{c}^2\sqrt{|\bar{p}|}+
 H_{\rm matter}(\bar{p})=0\,.
\end{equation}

Similarly, in the LRS Bianchi I model we have
\begin{eqnarray}
 A_a^i \md x^a\tau_i &=& A \tau^1 dx + A \tau^2 dy + c \tau^3 dz\\
 E^a_i \frac{\partial}{\partial x^a}\tau^i &=& p_A \tau^1 \partial_x +
p_A \tau^2 \partial_y + p_c
 \tau^3 \partial_z
\end{eqnarray}
with symplectic structure
\begin{eqnarray}
 \{A,\ p_A\} = 4\pi\gamma G \ , \quad \{c,\ p_c\} = 8\pi\gamma G \ .
\end{eqnarray}
There is a residual gauge transformation $p_A\mapsto-p_A$ which can be
fixed by requiring $p_A\geq0$ (see \cite{HomCosmo,BHInt} for more details).
The volume is now given by
\begin{equation}
 V_{\rm aniso} = \sqrt{p_A^2|p_c|}
\end{equation}
and the Hamiltonian constraint by
\begin{equation}
 H_{\rm aniso}= -(8\pi G)^{-1}\gamma^{-2}(A^2\sqrt{p_A}+ 2Ac\sqrt{|p_c|})+
H_{\rm matter}(p_A,p_c)=0\,.
\end{equation}

To identify the anisotropy as a perturbation it is convenient
to introduce new variables
\begin{eqnarray}
 (A,c) &=&
 (\bar{c}+\varepsilon,\ \bar{c}-2\varepsilon) \\
 (p_A,p_c) &=& (\bar{p}+ p_{\varepsilon},\ \bar{p}-2 p_{\varepsilon})
 \label{pApc}
\end{eqnarray}
together with the inverse transformation
\begin{eqnarray} \label{eq:cbareps}
 (\bar{c}, \bar{p}) &=& \left( {\textstyle\frac{1}{3}}(2A+c),
 {\textstyle\frac{1}{3}}(2p_A+p_C) \right) \\
 (\varepsilon, p_{\varepsilon}) &=&
 \left({\textstyle\frac{1}{3}}(A-c),{\textstyle\frac{1}{3}}(p_A-p_C)\right)
\end{eqnarray}
as well as the symplectic structure
\begin{eqnarray} \{\bar{c},\
 \bar{p} \} = \frac{8\pi}{3} \gamma G\ , \quad 
\{\varepsilon,\ p_{\varepsilon}\} =
 \frac{4\pi}{3} \gamma G
\ .
\end{eqnarray}
The volume, in this manner, is given by $\bar{p}$ up to terms of at
least second order in $p_{\varepsilon}$, which is the motivation for
choosing our particular set of anisotropy variables:
\begin{equation}
 V_{\rm aniso}= \sqrt{|\bar{p}^3-3p_{\varepsilon}^2\bar{p}-
2p_{\varepsilon}^3|}= |\bar{p}|^{3/2}\left(1-\frac{3}{2}
\frac{p_{\varepsilon}^2}{\bar{p}^2}+ O(p_{\varepsilon}^3/\bar{p}^3)\right)\,.
\end{equation}
This shows that we have to assume $p_{\varepsilon}\ll\bar{p}$ for
perturbations, i.e.\ the approximation will break down close to
classical singularities of the isotropic type where $\bar{p}=0$.
Later, we will also have to assume $\varepsilon\ll 1$.  For the
Hamiltonian constraint we obtain
\begin{equation}
 H_{\rm aniso}= -\frac{3}{8\pi G\gamma^2}|\bar{p}|^{-3/2}
 (\bar{c}^2\bar{p}^2-
\varepsilon^2\bar{p}^2- \bar{c}^2p_{\varepsilon}^2+2
\bar{c}\bar{p}\varepsilon p_{\varepsilon}+O(p_{\varepsilon}^3/\bar{p}^3))+
 H_{\rm matter}(\bar{p},p_{\varepsilon})
\end{equation}
up to terms of third order in the perturbation.

In the vacuum case, we have Hamiltonian equations of motion
\begin{eqnarray}
 \dot{\bar{c}} &=& \{\bar{c},H_{\rm aniso}\} \propto
-\bar{c}^2\bar{p}+\varepsilon^2\bar{p}-\bar{c}\varepsilon p_{\varepsilon}\\
 \dot{\bar{p}} &=& \{\bar{p},H_{\rm aniso}\} \propto
 \bar{c}\bar{p}^2-\bar{c}p_{\varepsilon}^2+\bar{p}\varepsilon
p_{\varepsilon}\\
 \dot{\varepsilon} &=& \{\varepsilon,H_{\rm aniso}\} \propto
 {\textstyle\frac{1}{2}}\bar{c}^2p_{\varepsilon}- {\textstyle\frac{1}{2}}
\bar{c}\bar{p}\varepsilon\\
 \dot{p}_{\varepsilon} &=& \{p_{\varepsilon},H_{\rm aniso}\} \propto
 -{\textstyle\frac{1}{2}} \bar{p}^2\varepsilon+ {\textstyle\frac{1}{2}}
\bar{c}\bar{p} p_{\varepsilon}\,.
\end{eqnarray}
Thus, the isotropic variables receive corrections in their equations
of motion only at second order. Ignoring those corrections, we recover
the classical solution $\bar{c}=0$ and $\bar{p}={\rm const}$ for flat
Minkowski space. With this approximate solution for the anisotropic
system, the equations for anisotropies simplify to
$\dot{\varepsilon}\approx0$ and $\dot{p}_{\varepsilon}\approx
-\frac{1}{2}\bar{p}^2\varepsilon$. The anisotropy in the connection will
thus remain constant while that in the triad grows linearly in time
until the approximations break down. Since the constraint is quadratic
in connection components, a feature shared by the full constraint, we
are not required at this point to assume $\varepsilon\ll 1$. However, the
equations of motion show that $\dot{p}_{\varepsilon}$ would be large if
$\varepsilon$ is not small, and we are led to this condition if the
perturbative evolution is to be valid for some time.


We will later, at the quantum level, mainly consider an asymptotic
approximation of very large $\bar{p}$ where all corrections of higher
order in $p_{\varepsilon}/\bar{p}$ are ignored. This also simplifies the
classical equations because the Hamiltonian constraint is then simply
proportional to $-(\bar{c}^2-\varepsilon^2)\sqrt{|\bar{p}|}$. Notice that
the quadratic nature of the constraint in connection variables makes
this approximation reasonable as we do not need to expand in
$\varepsilon$. The solution $\varepsilon=\bar{c}$ to the constraint is thus
consistent even though the anisotropy $\varepsilon$ in the connection is
not small compared to $\bar{c}$ (unlike $p_{\varepsilon}$ which must be
small compared to $\bar{p}$). As already mentioned, and used later, it
is only necessary that $\varepsilon$ is small compared to one, which is
usually the case also for $\bar{c}$ in semiclassical regimes.

Using the large-$\bar{p}$ constraint we have equations of motion
\begin{eqnarray*}
 \dot{\bar{c}} &=& -\frac{\bar{c}^2}{2\sqrt{|\bar{p}|}}+
 \frac{\varepsilon^2}{2\sqrt{|\bar{p}|}}\approx 0 \\
 \dot{\bar{p}} &=& 2\bar{c}\sqrt{|\bar{p}|}\\
 \dot{\varepsilon} &=& 0\\
 \dot{p}_{\varepsilon} &=& -\varepsilon\sqrt{|\bar{p}|}\,.
\end{eqnarray*}
With $\varepsilon=\bar{c}={\rm const}$ we have $\bar{p}=(\varepsilon
t+\sqrt{|\bar{p}_0|})^2$ and $p_{\varepsilon}= p_{\varepsilon}^0-
\frac{1}{2} \varepsilon^2t^2-\varepsilon \sqrt{|\bar{p}_0|}t$ with constants
of integration $\bar{p}_0$ and $p_{\varepsilon}^0$. Eliminating $t$, we have
the internal time evolution
\begin{equation} \label{pinternal}
 p_{\varepsilon}(\bar{p})=
 p_{\varepsilon}^0-{\textstyle\frac{1}{2}}(\bar{p}-\bar{p}_0)
\end{equation}
of $p_{\varepsilon}$ with respect to $\bar{p}$. Notice that this
evolution, unlike the coordinate time evolution, does not depend on
the value of $\varepsilon$. We will use this fact later as a test of the
perturbative quantum evolution.

\section{Quantization}

Loop quantizations are based on holonomies, and also in models one
uses exponentials of connection components as basic expressions in
addition to densitized triad components. In the isotropic case, this
leads to a basic algebra given by $p$ and $\exp(i\mu \bar{c}/2)$ for
all $\mu\in{\mathbb R}$, represented on a non-separable Hilbert space
\begin{equation}
 {\cal H}_{\rm iso}\cong
L^2(\bar{\mathbb R}_{\rm Bohr},{\rm d}\mu_{\rm H})
\end{equation}
of square integrable functions on the Bohr compactification of the
real line \cite{Bohr}, with orthonormal basis
\begin{equation}
 \langle \bar{c}|\mu\rangle = \exp(i\mu\bar{c}/2) \quad,\quad
\mu\in{\mathbb R}
\end{equation}
and basic operators acting as
\begin{eqnarray}
 \widehat{\exp(i\mu'\bar{c}/2)}|\mu\rangle &=& |\mu+\mu'\rangle 
\label{IsoHol}\\
  \hat{\bar{p}}|\mu\rangle &=&
{\textstyle\frac{1}{6}}\gamma\ell_{\rm P}^2\mu|\mu\rangle
\label{IsoFlux}
\end{eqnarray}
using the Planck length $\ell_{\rm P}=\sqrt{8\pi G\hbar}$.  Similarly,
for our anisotropic model we have the Hilbert space
\begin{equation}
{\cal H}_{\rm aniso}\cong {\cal H}_{\rm
iso}\otimes {\cal H}_{\rm iso}
\end{equation}
with orthonormal basis
\begin{equation}
 \langle A,c|\mu,\nu\rangle = \exp(i(\mu c+\nu A)/2) \quad,\quad
\mu,\nu\in{\mathbb R}
\end{equation}
holonomy operators as before and
\begin{eqnarray}
\hat{p}_A |\mu,\nu\rangle = {\textstyle\frac{1}{4}} \gamma\ell_{\rm P}^2 
\mu |\mu,\nu\rangle
\ , \quad \hat{p}_c |\mu,\nu\rangle = {\textstyle\frac{1}{2}}
 \gamma\ell_{\rm P}^2 \nu |\mu,\nu\rangle \,.
\end{eqnarray}
These states are not completely gauge invariant in view of the residual
gauge transformation $\mu\mapsto -\mu$, which however will not be
relevant for our purposes. One can always ensure gauge invariance by
working only with states $\sum_{\mu,\nu}\psi_{\mu,\nu}|\mu,\nu\rangle$
where $\psi_{\mu,\nu}$ is symmetric in $\mu$.
The triad operators then give us directly the volume operators for both
models by inserting them into the classical expressions.

For mathematical constructions one often makes use of dense subspaces
${\rm Cyl}_{\rm iso}$ and ${\rm Cyl}_{\rm aniso}$ of the two
Hilbert spaces given by the so-called cylindrical functions which are
finite linear combinations of the basis states $|\mu\rangle$ or
$|\mu,\nu\rangle$, respectively. With their algebraic duals of
distributional states (linear functionals on the cylindrical
subspaces) one has two Gel'fand triples ${\rm Cyl}_{\rm
iso}\subset {\cal H}_{\rm iso}\subset {\rm Cyl}_{\rm
iso}^{\star}$ and ${\rm Cyl}_{\rm aniso}\subset {\cal H}_{\rm
aniso}\subset {\rm Cyl}_{\rm aniso}^{\star}$ (see also \cite{ALMMT}).

\subsection{Symmetry reduction}

Since we are interested in the relation between a symmetric model and
a less symmetric one we first use the opportunity to demonstrate how a
reduced model can be obtained within quantum theory (see also
\cite{SymmRed,SphSymm,LivRev}). We start with the anisotropic model and,
following general constructions \cite{SymmRed}, define an isotropic
state as a distribution which is supported only on isotropic
connections, i.e.\ $\varepsilon=0$. To fulfill the definition, we must
find an antilinear map
\begin{equation} 
\sigma\colon{\rm Cyl}_{\rm
iso}\to {\rm Cyl}_{\rm aniso}^{\star}, |\mu\rangle \mapsto
(\mu|
\end{equation}
embedding the space of isotropic cylindrical functions into the
distributional dual of the anisotropic one such that
\[
 \sigma(|\mu\rangle)[|\rho,\tau\rangle]=
 \langle\mu|\rho,\tau\rangle|_{A=c} \quad \mbox{ for all }\ 
|\rho,\tau\rangle\,.
\]
On the right hand side, we use the restriction of a state
$|\rho,\tau\rangle$ to the subspace $A=c$ upon which it can be
interpreted as an isotropic state. Expanding $\sigma(|\mu\rangle)=:
\sum_{\kappa,\lambda}\sigma_{\kappa,\lambda}(\mu) \langle\kappa,\lambda|$,
we have
\[
\sigma_{\rho,\tau}(\mu)= \sum_{\kappa,\lambda}\sigma_{\kappa,\lambda}(\mu)
\delta_{\kappa,\rho} \delta_{\lambda,\tau}=
\sigma(|\mu\rangle)[|\rho,\tau\rangle]=
\int e^{-i\mu c/2}e^{i(\rho+\tau)c/2}{\rm d}\mu_{\rm H}(c)=
  \delta_{\mu,\rho+\tau}
\]
and thus
\begin{equation}
 (\mu|=\sigma(|\mu\rangle)= \sum_{\kappa,\lambda}\delta_{\mu,\kappa+\lambda}
 \langle\kappa,\lambda|= \sum_{\kappa}\langle\kappa,\mu-\kappa|\,.
\end{equation}
The summation is done over all real numbers, which is well-defined as
a distribution in ${\rm Cyl}_{\rm aniso}^{\star}$.

In fact, in this case the definition simply amounts to multiplying an
isotropic state in the connection representation with a
$\delta$-distribution supported at $\varepsilon=0$,
\[
 \sigma(|\mu\rangle)|A,c\rangle= \sum_{\kappa}e^{i(\kappa
   A+(\mu-\kappa)c)/2}= e^{i\mu c/2}\sum_{\kappa}e^{i\kappa(A-c)/2}=
 e^{i\mu c/2}\delta(A-c)
\]
but this would be more complicated in more general systems and the
full theory where the definitions of \cite{SymmRed} still apply.

Indeed, this gives $(\mu|\rho,\tau\rangle =
\delta_{\mu,\rho+\tau}$, i.e.\ the evaluation of the distribution is
non-zero only if the averaged label equals the isotropic one,
\[
 {\textstyle\frac{1}{2}}\gamma\ell_{\rm P}^2(\rho+\tau)=
 2(\hat{p}_A)_{\rho}+(\hat{p}_c)_{\tau}=3(\hat{\bar{p}})_{\rho+\tau}
\]
and $(\mu|h^{(\rho)}_Ah_c^{(\rho)-1}=(\mu|$, i.e.\ 
$\exp(i\rho\varepsilon)$ acts trivially with its dual action on symmetric
distributions which are thus supported only on isotropic connections
$\varepsilon=0$.

Here, we already used the dual action of anisotropic operators on
isotropic distributional states, which in this example mapped a
symmetric state to a symmetric one. This is not the case for arbitrary
operators, and one cannot simply define all operators for the reduced
model by the dual action of anisotropic operators \cite{SphSymm}.
However, one can do this for special operators which suffice to define
the basic operators of the model. For flux operators, one can easily
see that $2\hat{p}_A+\hat{p}_c=3\hat{\bar{p}}$
is the only one which maps an isotropic state $(\mu|$ to an isotropic
state:
\begin{equation}
 \hat{p}\sigma(|\mu\rangle)= {\textstyle\frac{1}{3}}(\mu|
   (2\hat{p}_A+\hat{p}_c)= {\textstyle\frac{1}{6}}\gamma\ell_{\rm P}^2
   \mu(\mu|= \sigma(\hat{\bar{p}}|\mu\rangle)\,.
\end{equation}
Moreover, it agrees with the isotropic flux operator (\ref{IsoFlux})
defined in the isotropic model. For holonomies, one can act with
arbitrary products of $h_A$ and $h_c$ on isotropic states since we
already know that $h_Ah_c^{-1}$ acts as the identity and the remaining
factor simply amounts to an isotropic holonomy operator. Also, all
holonomy operators form a closed algebra with the isotropic flux
operator $\hat{\bar{p}}=\frac{1}{3}(2\hat{p}_A+\hat{p}_c)$:
\begin{equation} \label{commp}
 [h_A^{(\rho)}h_c^{(\tau)},\hat{\bar{p}}] |\mu,\nu\rangle=
 -{\textstyle\frac{1}{6}}\gamma\ell_{\rm P}^2
 (\rho+\tau)|\mu+\rho,\nu+\tau\rangle=
 -{\textstyle\frac{1}{6}}\gamma\ell_{\rm P}^2 (\rho+\tau) 
 h_A^{(\rho)}h_c^{(\tau)}
 |\mu,\nu\rangle\,.
\end{equation}
This is similar for the anisotropy operator which we can define by
$\hat{p}_{\varepsilon}:=\frac{1}{3}(\hat{p}_A-\hat{p}_c)$. Here, we have
the commutator
\begin{equation}\label{commeps}
 [h_A^{(\rho)}h_c^{(\tau)},\hat{p}_{\varepsilon}]=
 -{\textstyle\frac{1}{6}}\gamma\ell_{\rm P}^2 
 ({\textstyle\frac{1}{2}}\rho-\tau) 
 h_A^{(\rho)}h_c^{(\tau)}\,.
\end{equation}
However, $\hat{p}_{\varepsilon}$ does not map an isotropic distribution
to another such distribution.

If we are looking for a reduction of the operator algebra to implement
isotropy, we need to find distinguished dual actions making use of the
form of symmetric states. One clearly distinguished operator is
$\hat{\bar{p}}$ because it is the only flux operator mapping an
isotropic distributional state to another such state. For holonomies,
$h_Ah_c^{-1}$ is distinguished because it fixes any isotropic
distributional state. Beyond that, no other operators are
intrinsically distinguished, but we do not yet have a useful algebra
of operators for an isotropic model: while $\bar{p}$ is available, our
distinguished holonomy corresponds to the anisotropy $\varepsilon$ rather
than the conjugate of $\bar{p}$.  Nevertheless, by acting with the
distinguished $\hat{\bar{p}}$ as well as {\em all} holonomy operators,
we can generate the isotropic representation. More precisely, we
define
\begin{equation}
 {\rm Stab}(\sigma):=\{\hat{O}\in {\cal A}_{\rm aniso}|\hat{O}
 \sigma({\rm Cyl}_{\rm iso})\subset \sigma({\rm Cyl}_{\rm
     iso})\}
\end{equation}
and
\begin{equation}
 {\rm Fix}(\sigma):=\{\hat{O}\in {\cal A}_{\rm aniso}|
 \hat{O}\sigma(\psi)=\sigma(\psi) \mbox{ for all
 }\psi \in {\rm Cyl}_{\rm iso}\}\,.
\end{equation}
The isotropic algebra is then
\begin{equation}
 {\cal A}_{\rm iso}:={\rm Stab}(\sigma)/{\rm Fix}(\sigma)
\end{equation}
understood in the following manner: ${\rm Fix}(\sigma)$ contains only
holonomy operators besides the identity, and thus forms a normal
subgroup of the Abelian group obtained from the multiplicative
operation in the holonomy algebra. We can thus take the factor group
of holonomy operators modulo ${\rm Fix}(\sigma)$ which thanks to the
distributive law can also be equipped with the usual additive
structure to form an algebra.  Due to the relation (\ref{commp}), also
the flux operators in ${\rm Stab}(\sigma)$ commute with ${\rm
Fix}(\sigma)$ and thus descend to the factor group, defining ${\cal
A}_{\rm iso}$. Using that ${\rm Stab}(\sigma)$ is generated as an
algebra by $\hat{\bar{p}}$ and all $h_A^{(\rho)}h_c^{(\tau)}$ and
${\rm Fix}(\sigma)$ by $h_Ah_c^{-1}$, a general element of ${\cal
A}_{\rm iso}$ is
\[
 \left(\sum_{n\in{\mathbb N}_0} \hat{\bar{p}}^n \pi_n(h_c)\right)
\cdot {\rm Fix}(\sigma)
\]
with Laurent polynomials $\pi_n$, allowing also inverse powers of
$h_c$, for each $n$ which are non-zero only for finitely many $n$ (we
can order $\hat{p}$ to the left using (\ref{commp}) and factor out a
power of $h_Ah_c^{-1}$ into ${\rm Fix}(\sigma)$ such that only the
holonomy $h_c$ appears explicitly).  Through this construction we
derive an algebra isomorphic to the isotropic one.

As the presence of factor spaces indicates, there is no canonical
isotropic subalgebra of the anisotropic algebra without additional
input. This is analogous to the classical situation where the
splitting of anisotropic variables $A$ and $c$ into an isotropic
average $\bar{c}$ and an anisotropy is not unique.  Classically, a
given $\bar{p}$, as we have it distinguished at the quantum level,
determines a form of $\varepsilon$ as a linear combination of $A$ and $c$
by requiring $\{\bar{p},\varepsilon\}=0$. Indeed, also the quantum analog
of $\varepsilon$, given by $h_Ah_c^{-1}$, is distinguished as it commutes
with $\hat{\bar{p}}$. However, a unique form for $\bar{c}$ and
$p_{\varepsilon}$ then follows only after an additional choice, which for
us was the fact that the volume should receive corrections only to
second order. We use this here to define
$\hat{p}_{\varepsilon}=\frac{1}{3}(\hat{p}_A-\hat{p}_c)$ as above and
then require that (\ref{commeps}) vanish, analogously to
$\{\bar{c},p_{\varepsilon}\}=0$.  This distinguishes $h_A^2h_c$ as the
isotropic holonomy operator in addition to the flux $\hat{\bar{p}}$
which indeed form a subalgebra of the anisotropic operator algebra
mapping isotropic states to isotropic states in a way isomorphic to
the isotropic model.

In this manner, we obtain a unique subalgebra of the basic anisotropic
operator algebra which is isomorphic to the basic isotropic algebra.
This also presents an alternative way to define a symmetric model
without referring to symmetric distributional states at all: we use
the classical relations between symmetric and non-symmetric variables
to define a distinguished algebra derived from the non-symmetric
holonomy-flux algebra.  Using the fact that the non-symmetric
representation is cyclic on $|0,0\rangle$, i.e.\ the subspace ${\cal
A}_{\rm aniso} |0,0\rangle$ is dense in ${\cal H}_{\rm aniso}$, we
generate the representation of the reduced model by acting only with
the subalgebra on the cyclic state. This defines the reduced state
space, which is equipped with an inner product by requiring holonomy
operators to be unitary. Upon completion, this inner product space
defines the Hilbert space of the isotropic model and its quantum
representation in agreement with the loop quantization of the
classically reduced model. Characteristic properties of the
representation are then clearly inherited from that of the less
symmetric system, which in some cases is also visible for the
reduction from the full theory
\cite{SphSymm,LivRev}.

\subsection{Perturbations}

Dynamical information is obtained by quantizing the Hamiltonian
constraints and asking physical states to be annihilated by the
resulting operators. Because the constraints contain connection
components, and connections are represented via holonomies which shift
the labels of states as in (\ref{IsoHol}), quantum constraint
equations become difference equations
\cite{cosmoIII,IsoCosmo,HomCosmo} for states $|\psi\rangle =
\sum_{\mu}\psi_{\mu}|\mu\rangle$ or $|\psi\rangle =
\sum_{\mu,\nu}\psi_{\mu,\nu}|\mu,\nu\rangle$ in the triad
representation.  For large scales, e.g.\ $\mu\gg1$ in the isotropic
case, the difference equation is well-approximated by the
Wheeler--DeWitt equation \cite{SemiClass} while for small $\mu$ there
are crucial differences implying the absence of singularities at the
quantum level \cite{Sing}. In order to see the relation to the
Wheeler--DeWitt equation, one has to assume that the wave function
does not vary rapidly on small scales, i.e.\ changes in $\mu$ of the
order one, which allows one to Taylor expand the difference operators
into differential ones. This limit, however, does not exist for
arbitrary states because there is no operator for $\bar{c}$ on ${\cal
H}_{\rm iso}$ \cite{Bohr}.  Indeed, the Wheeler--DeWitt quantization
is based on a representation on the usual Schr\"odinger Hilbert space
${\cal H}_{\rm S}=L^2({\mathbb R},\md \bar{c})$ which is inequivalent
to the representation on the Bohr Hilbert space used for the loop
quantization.

In this paper, we analyze the case of an anisotropic model where,
however, difference operators are not expanded for both triad
components but only for that corresponding to the anisotropy
$p_{\varepsilon}$. We thus assume that the wave function does not vary
rapidly when $p_{\varepsilon}$ changes and that in an approximate
sense an operator for $\varepsilon$ exists. Although we need to assume
$p_{\varepsilon}$ to be small, this will not push us in the regime
where discreteness is relevant provided that $\bar{p}$ is large. The
key additional assumption is that $\varepsilon$ is not too big which
now becomes important at the quantum level, and indeed both anisotropy
parameters, $p_{\varepsilon}$ and $\varepsilon$, have to be
small. (The situation is different from pure isotropy where small
$\bar{p}$ usually implies large $\bar{c}$.) When only small
$\varepsilon$ are allowed, we do not probe the whole configuration
space and do not see its compactness which underlies the Bohr Hilbert
space used in homogeneous loop quantizations.  As in the large volume
limit, we therefore introduce the Schr\"odinger Hilbert space ${\cal
H}_{\rm S}$ for anisotropies, which we do by defining the perturbative
Hilbert space
\begin{equation}
{\cal H}_{\rm pert}  := {\cal
  H}_{\rm iso} \otimes {\cal H}_{\rm S}
\end{equation}
and realizing its dense subset ${\rm Cyl}_{\rm iso}\otimes{\rm
Cyl}_{\rm S}$ as a subspace of the dual ${\rm Cyl}_{\rm
aniso}^{\star}$. In the Schr\"odinger Hilbert space we have to choose a
suitable dense set ${\rm Cyl}_{\rm S}$, which for our purposes will be
the set of all functions which are products of a polynomial and a
Gaussian. Indeed, Schr\"odinger states in ${\rm Cyl}_{\rm S}$ can be
interpreted as distributions on the Bohr Hilbert space
\cite{PolymerParticle} using the antilinear map
\[
 \pi\colon {\rm Cyl}_{\rm S}\to {\rm Cyl}_{\rm Bohr}^{\star},
 \pi(\psi)[|\rho\rangle]:=\int
 e^{i\rho\varepsilon}\overline{\psi(\varepsilon)}\md\varepsilon
\,.
\]
Dual actions of basic operators on those states are given by
\begin{eqnarray} \label{Dualp}
 (\hat{p}_{\varepsilon}\pi(\psi))[|\rho\rangle] &=&
 \pi(\psi)[\hat{p}_{\varepsilon}^{\dagger}|\rho\rangle]=
{\textstyle\frac{1}{6}\gamma\ell_{\rm P}^2}
\int \rho e^{i\rho\varepsilon}\overline{\psi(\varepsilon)}\md\varepsilon=
{\textstyle\frac{1}{6}\gamma\ell_{\rm P}^2}
\int e^{i\rho\varepsilon} \cdot i\frac{\md}{\md\varepsilon} 
\overline{\psi(\varepsilon)}\md\varepsilon\\
&=&\pi(-{\textstyle\frac{1}{6}i\gamma\ell_{\rm P}^2}
\md\psi/\md\varepsilon)[|\rho\rangle]\nonumber
\end{eqnarray}
and
\begin{equation} \label{Dualeps}
 (e^{i\tau\varepsilon}\pi(\psi))[|\rho\rangle]=
\psi[e^{-i\tau\varepsilon}|\rho\rangle]= \int e^{i(\rho-\tau)\varepsilon} 
\overline{\psi(\varepsilon)}\md\varepsilon=
\int e^{i\rho\varepsilon} 
\overline{e^{i\tau\varepsilon}\psi(\varepsilon)}\md\varepsilon
=\pi(e^{i\tau\varepsilon}\psi)[|\rho\rangle]\,.
\end{equation}
The first equation shows that the momentum operator is just the
derivative operator, as also on the Bohr Hilbert space, while the
second shows that $(e^{i\tau\varepsilon}\pi(\psi))(\varepsilon)=
e^{i\tau\varepsilon}\psi(\varepsilon)$ where on the Schr\"odinger Hilbert
space we can now take the derivative with respect to $\nu$. By going
to the dual action on the image of $\pi$, we thus obtain a simple
multiplication operator for $\varepsilon$ well-defined with domain
$\pi({\rm Cyl}_{\rm S})$.

In addition, if we choose an appropriate state $\Psi\in{\rm Cyl}_{\rm
S}$ we have an embedding ${\rm id}\otimes\Psi\colon {\rm Cyl}_{\rm
iso}\to {\rm Cyl}_{\rm pert}$ of isotropic cylindrical states to
perturbative cylindrical states where anisotropies are small in mean
value, but not eliminated exactly.  Moreover, since we have to choose
a state $\Psi$, which in the above notation is identified with the map
${\mathbb C}\to{\rm Cyl}_{\rm S},1\mapsto\Psi$, there is no unique
canonical embedding. (Coherent states are natural candidates, but
unlikely to be preserved dynamically.)  In addition to the strict
implementation of symmetries $\sigma$ we then have the perturbative
implementation $\pi_{\Psi}:= (\star\otimes\pi)\circ({\rm
id}\otimes\Psi)$ which are both maps from ${\rm Cyl}_{\rm iso}$ to
${\rm Cyl}_{\rm aniso}^{\star}$:
\begin{center}
\unitlength1mm
\begin{picture}(100,30)
 \put(5,25){\makebox(0,0){${\rm Cyl}_{\rm iso}^{\star}$}}
 \put(93,25){\makebox(0,0){${\rm Cyl}_{\rm aniso}^{\star}$}}
 \put(5,15){\makebox(0,0){${\cal H}_{\rm iso}$}}
 \put(93,15){\makebox(0,0){${\cal H}_{\rm aniso}$}}
 \put(5,5){\makebox(0,0){${\rm Cyl}_{\rm iso}$}}
 \put(93,5){\makebox(0,0){${\rm Cyl}_{\rm aniso}$}}
 \put(5,20){\makebox(0,0){$\bigcup$}}\put(93,20){\makebox(0,0){$\bigcup$}}
 \put(5,10){\makebox(0,0){$\bigcup$}}\put(93,10){\makebox(0,0){$\bigcup$}}
 \put(50,5){\makebox(0,0){${\rm Cyl}_{\rm iso}\otimes {\rm Cyl}_{\rm S}= 
{\rm Cyl}_{\rm pert}$}}
 \put(10,6.5){\vector(4,1){75}}
 \put(10,5.5){\vector(1,0){18}}
 \put(52,8){\vector(2,1){33}}
 \put(40,16){\makebox{$\sigma$}}
 \put(19,3){\makebox(0,0){${\rm id}\otimes\Psi$}}
 \put(70,14){\makebox{$\star\otimes\pi$}} 
\end{picture}
\end{center}
We have the two Gel'fand triples of the exact models on the left and
right, with the canonical embedding $\sigma$ of the reduced model
through distributions implementing symmetries precisely. In between,
we have the perturbative cylindrical space which is connected to the
anisotropic side through the antilinear map $\pi$ from ${\rm Cyl}_{\rm
S}$, combined with the antilinear dualization $\star$ of isotropic
states.

In the perturbative sector, non-symmetric degrees of freedom are
present but can be arranged to be unexcited at least for initial
states of the evolution. In this sense, the situation is similar to
the field theory example studied in \cite{SymmQFT} where non-symmetric
field modes are included in their vacuum or other coherent states. In
this case, the field theory vacuum is a natural candidate for the
analog of $\Psi$. Perturbative states can then be thought of as states
where only the symmetric modes have been excited but non-symmetric
ones have been left in their vacua. This is technically similar to our
construction of strictly symmetric states at the end of the preceding
subsection where we used only symmetric operators in order to generate
all states of the symmetric model out of the non-symmetric cyclic
state $|0,0\rangle$. Indeed, for background independent systems in
loop quantum gravity the analog of a ground state is the cyclic state
where no geometry at all is excited. Thus, although physically the
construction in \cite{SymmQFT} is closer to the perturbative situation
here, technically it is analogous to the situation of the strict
model. This illustrates crucial differences in the context of
symmetries between quantum field theory on a background and a
background independent quantization of gravity.

We have already discussed the relation between anisotropic operators
and isotropic ones for the exact implementation of symmetry. For the
perturbative implementation, the same relation exists for operators
which do not depend on anisotropies, but we can now also act with
other operators as we already saw for $\hat{\varepsilon}$ and
$\hat{p}_{\varepsilon}$ in (\ref{Dualeps}) and (\ref{Dualp}). In general,
an anisotropic operator $\hat{O}$ acting on ${\cal H}_{\rm aniso}$ has
a dual action on $\star\otimes\pi({\rm Cyl}_{\rm pert})$, but does not
necessarily fix this subspace of ${\rm Cyl}_{\rm
aniso}^{\star}$. However, it does so perturbatively when we expand it
as a sum of operators in the perturbative sector. Choosing, for
definiteness, a Gaussian for $\Psi$ we have perturbative states of the
form
\[
 \psi(A,c)= e^{i\bar{\nu}\bar{c}/2}
 e^{-(\varepsilon-\varepsilon_0)^2/4\sigma^2} e^{6i\varepsilon
p_{\varepsilon}^0/\gamma\ell_{\rm P}^2}
\]
where $\bar{c}=\frac{1}{3}(2A+c)$ and $\varepsilon=\frac{1}{3}(A-c)$ are
understood as functions of $A$ and $c$. By the chain rule, we then
have flux operators
\[
 \hat{p}_A\psi=-\frac{i}{2}\gamma\ell_{\rm P}^2\frac{\partial}{\partial A}\psi=
 -\frac{i}{6}\gamma\ell_{\rm P}^2 \left(2\frac{\partial}{\partial\bar{c}}+
   \frac{\partial}{\partial\varepsilon}\right)\psi=
 \left({\textstyle\frac{1}{6}}\gamma\ell_{\rm P}^2
   \bar{\nu}+\hat{p}_{\varepsilon}\right)\psi
\]
and
\[
\hat{p}_c\psi= -i\gamma\ell_{\rm P}^2\frac{\partial}{\partial c}\psi=
-\frac{i}{3}\gamma\ell_{\rm P}^2 \left(\frac{\partial}{\partial\bar{c}}-
  \frac{\partial}{\partial\varepsilon}\right)\psi=
\left({\textstyle\frac{1}{6}}\gamma\ell_{\rm P}^2\bar{\nu}-
  2\hat{p}_{\varepsilon}\right)\psi\,.
\]
In composite expressions such as the volume operator one can then
expand functions of flux operators in $p_{\varepsilon}$ when they are
expressed in the flux representation, and analogously for holonomy
operators in the connection representation. 

This is most easily done
if eigenvalues are already known. For instance, if an operator
$\hat{O}$ has eigenstates $|\mu,\nu\rangle$, we take the eigenvalues
$O_{\mu,\nu}$ and insert
 \begin{eqnarray} \label{mandn}
 \nonumber \mu&=&{\textstyle\frac{2}{3}}\bar{\nu}
 +4\gamma^{-1}\ell_{\rm P}^{-2}p_{\varepsilon}= 
 {\textstyle\frac{2}{3}}\bar{\nu}+P\,,\\
 \nu&=&{\textstyle\frac{1}{3}}\bar{\nu}
 -4\gamma^{-1}\ell_{\rm P}^{-2}p_{\varepsilon}=
 {\textstyle\frac{1}{3}}\bar{\nu}-P
\end{eqnarray}
with dimensionless $P:=4p_{\epsilon}/\gamma\ell_{\rm P}^2$, which
follows from the classical relation (\ref{pApc}) or the calculation
above. This yields a function $O(\bar{\nu},P)$ which we
expand in the perturbation $P$,
\begin{equation}
 O(\bar{\nu},P)=\sum_k O^{(k)}_{\rm iso}(\bar{\nu}) P^k \,.
\end{equation}
Note that $P$ itself need not be small by our assumptions, which would
mean $p_{\varepsilon}\ll\ell_{\rm P}^2$. We have, however,
$P\ll\bar{\nu}$ such that each $O^{(k)}_{\rm iso}(\bar{\nu})$ must
drop off at least as $\bar{\nu}^{-k}$.  For any fixed $k$, the values
$O^{(k)}_{\rm iso}(\bar{\nu})$, interpreted as eigenvalues, define an
isotropic operator $\hat{O}^{(k)}_{\rm iso}=\sum_{\bar{\nu}}
O^{(k)}_{\rm iso}(\bar{\nu})|\bar{\nu}\rangle \langle\bar{\nu}|$ such
that $\hat{O}^{(k)}_{\rm iso} |\bar{\nu} \rangle:= O^{(k)}_{\rm
iso}(\bar{\nu}) |\bar{\nu} \rangle$. Thus, we obtain the expansion
\begin{eqnarray} \label{OpExpand}
 \hat{O}\sim \sum_k \hat{O}^{(k)}_{\rm iso}\otimes \hat{P}^k
\end{eqnarray}
where the right hand side acts on ${\cal H}_{\rm pert}$. This
approximation of operators on a selected class of states implementing
small anisotropy is discussed in more detail in App.~\ref{ApproxOp},
which can be applied to any Hilbert space $|\bar{\nu}\rangle\otimes
{\cal H}_{\rm S}$ for fixed $\bar{\nu}$.  We understand here always a
finite truncation to a given order in $k$, and do not discuss
convergence of the expansion in whatever sense.
We will later use states $\Psi$ of the above form, assuming
$\sigma<\varepsilon_0\ll1$ in order to expand also in $\varepsilon$. This
results in similar expansions for operators containing holonomies,
such as the Hamiltonian constraint.

\subsection{Hamiltonian constraint} 

The Hamiltonian constraint operator for the anisotropic model results
from the expression given in \cite{HomCosmo} after using $A=c_1=c_2$
and $c=c_3$ (as in \cite{HomCosmo}, we ignore an ambiguity parameter
$\mu_0$ \cite{Bohr} which is of the order one and multiplies
connection components in the constraint):
\begin{eqnarray}
 \nonumber \hat{H}&=& 4\pi^{-1} i \gamma^{-3} G^{-1} \ell_{\rm P}^{-2}
 \left\{ \sin^2\left( {\textstyle\frac{1}{2}}A\right) 
\cos^2\left( {\textstyle\frac{1}{2}}A
 \right) \hat {O}_c + 2 \sin\left( {\textstyle\frac{1}{2}}c \right) \cos\left(
 {\textstyle\frac{1}{2}}c\right) \sin\left( 
{\textstyle\frac{1}{2}}A \right) \cos\left(
 {\textstyle\frac{1}{2}}A\right) \hat{O}_A \right\}\\
\hat {O}_c&:=&
 \sin\left( {\textstyle\frac{1}{2}}c \right) \hat{V} 
\cos\left( {\textstyle\frac{1}{2}}c\right) -
 \cos\left( {\textstyle\frac{1}{2}}c\right) \hat{V} 
\sin\left( {\textstyle\frac{1}{2}}c \right)
\end{eqnarray}
and similarly $\hat{O}_A$ by replacing $c$ by $A$,
where $\hat{V}$ is the anisotropic volume. Making use of the
change of variables (\ref{eq:cbareps}) one gets
\begin{eqnarray}
 \label{eq1} \hat{H} &=& 4\pi^{-1} i \gamma^{-3} G^{-1}
 \ell_{\rm P}^{-2} \left\{ \sin^2\left( {\textstyle\frac{1}{2}}\bar{c}+
 {\textstyle\frac{1}{2}}\varepsilon\right) \cos^2\left( {\textstyle\frac{1}{2}}\bar{c}+
 {\textstyle\frac{1}{2}}\varepsilon \right) \hat {O}_{c}
 \right. \nonumber \\ && \mbox{} + 2 \left. \sin\left(
 {\textstyle\frac{1}{2}}\bar{c}-\varepsilon \right) \cos\left(
 {\textstyle\frac{1}{2}}\bar{c}-\varepsilon\right) \sin\left( {\textstyle\frac{1}{2}}\bar{c}+
 {\textstyle\frac{1}{2}}\varepsilon \right) \cos\left( {\textstyle\frac{1}{2}}\bar{c}+
 {\textstyle\frac{1}{2}}\varepsilon\right) \hat{O}_{A}
 \right\}
\end{eqnarray}

Here $\hat {O}_{c}$ and $\hat{O}_{A}$ are more conveniently
dealt with by using their action on the $|\mu,\nu \rangle$ basis, namely,
\begin{eqnarray}
 \hat {O}_A |\mu,\nu \rangle &=& \frac{i}{2}
 \left(V_{\mu+1,\nu}-V_{\mu-1,\nu} \right) \ |\mu,\nu \rangle\\
 \hat {O}_c |\mu,\nu \rangle &=& \frac{i}{2}
 \left(V_{\mu,\nu+1}-V_{\mu,\nu-1} \right) \ |\mu,\nu \rangle
\end{eqnarray}
Now following (\ref{OpExpand}) we expand $\hat{O}_A$ and
$\hat{O}_c$ up to second order in $p_{\varepsilon}$, using
from the exact analysis \cite{HomCosmo} that
\begin{equation}
 \label{eq:volmn} V_{\mu, \nu} =
 \frac{1}{2}\left( \frac{1}{2}\gamma\ell_{\rm P}^2
 \right)^{\frac{3}{2}}\sqrt{\mu^2|\nu|}= 
\frac{1}{2}\left( \frac{1}{2}\gamma\ell_{\rm P}^2
 \right)^{\frac{3}{2}} |\mu| \sqrt{|\nu|}\,.
\end{equation}
Inserting (\ref{mandn}), we obtain the eigenvalues
\begin{eqnarray*}
(O_A)_{\mu(\bar{\nu},P),\nu(\bar{\nu},P)} &=& {\textstyle\frac{1}{4}}i \left( 
{\textstyle\frac{1}{2}}\gamma
 \ell_{\rm P}^2\right)^{3/2} \left(|{\textstyle\frac{2}{3}}\bar{\nu}+1+P|-
|{\textstyle\frac{2}{3}}\bar{\nu}-1+P|\right)
\sqrt{|{\textstyle\frac{1}{3}}\bar{\nu}-P|} \\
(O_c)_{\mu(\bar{\nu},P),\nu(\bar{\nu},P)} &=& {\textstyle\frac{1}{4}}i \left( 
{\textstyle\frac{1}{2}}\gamma
 \ell_{\rm P}^2\right)^{3/2}
|{\textstyle\frac{2}{3}}\bar{\nu}+P|
\left(\sqrt{|{\textstyle\frac{1}{3}}\bar{\nu}+1-P|}-
\sqrt{|{\textstyle\frac{1}{3}}\bar{\nu}-1-P|}\right) \,.
\end{eqnarray*}
For $P\ll\bar{\nu}$, the absolute values can be evaluated as
\begin{eqnarray*}
 |{\textstyle\frac{2}{3}}\bar{\nu}+1+P|-
|{\textstyle\frac{2}{3}}\bar{\nu}-1+P| &=&
|{\textstyle\frac{2}{3}}\bar{\nu}+1| \left(1+
({\textstyle\frac{2}{3}}\bar{\nu}+1)^{-1}P\right)-
|{\textstyle\frac{2}{3}}\bar{\nu}-1| \left(1+ 
({\textstyle\frac{2}{3}}\bar{\nu}-1)^{-1}P\right)\\
 &=& |{\textstyle\frac{2}{3}}\bar{\nu}+1|-
|{\textstyle\frac{2}{3}}\bar{\nu}-1|+
\left(\sig({\textstyle\frac{2}{3}}\bar{\nu}+1)-
\sig({\textstyle\frac{2}{3}}\bar{\nu}-1)\right)P\\
 &=& 2\sig_{3/2}(\bar{\nu})- 3\chi_{3/2}(\bar{\nu})P\,,
\end{eqnarray*}
introducing
\begin{eqnarray}
 \chi_{\delta}(\bar{\nu}) &:=& -\frac{1}{2\delta}
(\sig(\bar{\nu}+\delta)- \sig(\bar{\nu}-\delta)) = \left\{
\begin{array}{cl} 0 & \mbox{for }|\bar{\nu}|\geq\delta\\ \delta^{-1} &
\mbox{for }|\bar{\nu}|<\delta \end{array}\right.\\
\sig_{\delta}(\bar{\nu}) &:=& \frac{1}{2\delta}(|\bar{\nu}+\delta|-
|\bar{\nu}-\delta|) =  \left\{
\begin{array}{cl} 1 & \mbox{for }\bar{\nu}\geq\delta\\ \delta^{-1}\bar{\nu} &
\mbox{for }|\bar{\nu}|<\delta\\ -1 &  \mbox{for }\bar{\nu}\leq\delta 
\end{array}\right.\,,
\end{eqnarray}
and
\[
 |{\textstyle\frac{2}{3}}\bar{\nu}+P|=
|{\textstyle\frac{2}{3}}\bar{\nu}| (1+3P/2\bar{\nu})
\]
while the other expressions are expanded as
\[
 \sqrt{|{\textstyle\frac{1}{3}}\bar{\nu}-P|} =
\sqrt{{\textstyle\frac{1}{3}}\bar{\nu}}
\left(1-{\textstyle\frac{3}{2}}P/\bar{\nu}-
{\textstyle\frac{9}{8}}P^2/\bar{\nu}^2+ O(P^3/\bar{\nu}^3)\right)
\]
and
\begin{eqnarray*}
 |{\textstyle\frac{1}{3}}\bar{\nu}+1-P|^{1/2}-
|{\textstyle\frac{1}{3}}\bar{\nu}-1-P|^{1/2} &=&
|{\textstyle\frac{1}{3}}\bar{\nu}+1|^{1/2}
\sqrt{1-({\textstyle\frac{1}{3}}\bar{\nu}+1)^{-1}P}\\
&&-
|{\textstyle\frac{1}{3}}\bar{\nu}-1|^{1/2}
\sqrt{1-({\textstyle\frac{1}{3}}\bar{\nu}-1)^{-1}P}\\
&=& |{\textstyle\frac{1}{3}}\bar{\nu}+1|^{1/2}
(1-{\textstyle\frac{1}{2}} ({\textstyle\frac{1}{3}}\bar{\nu}+1)^{-1}P-
{\textstyle\frac{1}{8}}
({\textstyle\frac{1}{3}}\bar{\nu}+1)^{-2}P^2)\\
&&-
|{\textstyle\frac{1}{3}}\bar{\nu}-1|^{1/2} (1-{\textstyle\frac{1}{2}}
({\textstyle\frac{1}{3}}\bar{\nu}-1)^{-1}P- 
{\textstyle\frac{1}{8}} ({\textstyle\frac{1}{3}}\bar{\nu}-1)^{-2}P^2)\\
&&+O(P^3/\bar{\nu}^3)\\
&=& |{\textstyle\frac{1}{3}}\bar{\nu}+1|^{1/2}-
|{\textstyle\frac{1}{3}}\bar{\nu}-1|^{1/2}\\
&&- {\textstyle\frac{1}{2}}
(|{\textstyle\frac{1}{3}}\bar{\nu}+1|^{-1/2}
\sig({\textstyle\frac{1}{3}}\bar{\nu}+1)-
|{\textstyle\frac{1}{3}}\bar{\nu}-1|^{-1/2}
\sig({\textstyle\frac{1}{3}}\bar{\nu}-1))P\\
&&- {\textstyle\frac{1}{8}}
(|{\textstyle\frac{1}{3}}\bar{\nu}+1|^{-3/2}-
|{\textstyle\frac{1}{3}}\bar{\nu}-1|^{-1/2})P^2+ O(P^3/\bar{\nu}^3)\\
&=& 2\sqrt{3} \Delta_3\sqrt{|\bar{\nu}|}-
3\sqrt{3}P\Delta_3\frac{\sig(\bar{\nu})}{\sqrt{|\bar{\nu}|}}\\
&&-
{\textstyle\frac{9}{4}}\sqrt{3}P^2\Delta_3|\bar{\nu}|^{-3/2}+
O(P^3/\bar{\nu}^3)
\end{eqnarray*}
introducing, more generally than before,
\begin{equation}
 \Delta_{\delta}f(\bar{\nu}) := \frac{1}{2\delta}(f(\bar{\nu}+\delta)-
f(\bar{\nu}-\delta))
\end{equation}
for any function $f$ which for functions differentiable at large
$\bar{\nu}$ gives $\Delta_{\delta}f(\bar{\nu})\sim f'(\bar{\nu})$ for
$\bar{\nu}\gg\delta$.

This results in
\begin{eqnarray} 
 \label{eq:omna} \nonumber \hat{O}_A&=& {\textstyle\frac{1}{4\sqrt{3}}}i\left( 
{\textstyle\frac{1}{2}}\gamma
 \ell_{\rm P}^2\right)^{3/2} \sqrt{|\hat{\bar{\nu}}|}
\left(2\sig_{3/2}(\hat{\bar{\nu}})\otimes{\rm id} -
3(\chi_{3/2}(\hat{\bar{\nu}}) +
\hat{\bar{\nu}}^{-1}\sig_{3/2}(\hat{\bar{\nu}})) \otimes \hat{P}
\right.\nonumber\\
&&+ 
\left.{\textstyle\frac{9}{2}} \hat{\bar{\nu}}^{-1}
(\chi_{3/2}(\hat{\bar{\nu}})- {\textstyle\frac{1}{2}}
\hat{\bar{\nu}}^{-1} \sig_{3/2}(\hat{\bar{\nu}})) \otimes \hat{P}^2+
O(P^3/\bar{\nu}^3)\right)
\end{eqnarray}
and
\begin{eqnarray} \label{eq:omnc}
 \nonumber \hat{O}_c&=& {\textstyle\frac{1}{2\sqrt{3}}}i\left( 
{\textstyle\frac{1}{2}}\gamma
 \ell_{\rm P}^2\right)^{3/2} |\hat{\bar{\nu}}|
\left(2\Delta_3|\hat{\bar{\nu}}|^{1/2} \otimes {\rm id} +
3(\hat{\bar{\nu}}^{-1} \Delta_3 |\hat{\bar{\nu}}|^{1/2}-
\Delta_3(|\hat{\bar{\nu}}|^{-1/2}\sig(\hat{\bar{\nu}}))) \otimes
\hat{P}\right.\nonumber\\
&&- \left.{\textstyle\frac{9}{2}} (\hat{\bar{\nu}}^{-1}
\Delta_3(|\hat{\bar{\nu}}|^{-1/2}\sig(\hat{\bar{\nu}}))+
{\textstyle\frac{1}{2}} \Delta_3|\hat{\bar{\nu}}|^{-3/2}) \otimes
\hat{P}^2 +O(P^3/\bar{\nu}^3)\right)\,.
\end{eqnarray}
where $\hat{\bar{\nu}}= 6\hat{\bar{p}}/(\gamma\ell_{\rm P}^2)$ and
$\hat{P}=4\hat{p}_{\varepsilon}/(\gamma \ell_{\rm P}^2)$. 
{}From these expressions we can read off the first coefficients
$O_{A,{\rm iso}}^{(k)}$ and $O_{c,{\rm iso}}^{(k)}$.


Note that, while the expressions $\chi_{\delta}$, $\sig_{\delta}$ and
$\Delta_{\delta}$ can be directly extended to be applied to operators,
the expansions are not densely defined on ${\cal H}_{\rm pert}$
because of inverse powers of $\hat{\bar{\nu}}$. In fundamental
expressions where the classical analog has inverse powers of
$\hat{\bar{p}}$, such as matter Hamiltonians, one can still define
densely defined operators
\cite{QSDV,InvScale}. Here, however, we started with densely defined
operators and obtained inverse powers only after expanding in the
perturbative sector. We thus cannot change those terms and make them
densely defined. For what follows we can mostly ignore this issue,
disregarding states at small $\bar{\nu}$ on which the operators cannot
be applied.  We will come back to it in the discussion of
singularities in Sec.~\ref{s:Sing}.

Next we must have the action of the several terms to the left of
$\hat{O}_A$ and $\hat{O}_c$ in ($\ref{eq1}$). Up to
order $\varepsilon^3$ we have
\begin{eqnarray}\label{eqH}
\nonumber \hat{H}&=& 4\pi^{-1} i \gamma^{-3} G^{-1}
\ell_{\rm P}^{-2}\bigg[2\left\{\sin^2{{\textstyle\frac{1}{2}}\bar{c}}
\cos^2{{\textstyle\frac{1}{2}}\bar{c}}\otimes{\rm id}+
{\textstyle\frac{1}{2}}\left(-\sin{{\textstyle\frac{1}{2}}\bar{c}}
\cos^3{{\textstyle\frac{1}{2}}\bar{c}}+\sin^3{{\textstyle\frac{1}{2}}
\bar{c}}\cos{{\textstyle\frac{1}{2}}\bar{c}}
\right)\otimes\varepsilon \right. \\\nonumber
 && +{\textstyle\frac{1}{2}}\left(-3\sin^2{{\textstyle\frac{1}{2}}\bar{c}}
\cos^2{{\textstyle\frac{1}{2}}\bar{c}}-
\cos^4{{\textstyle\frac{1}{2}}\bar{c}}-\sin^4{{\textstyle\frac{1}{2}}\bar{c}}
\right)\otimes\varepsilon^2+
{\textstyle\frac{1}{12}}\left(\sin{{\textstyle\frac{1}{2}}\bar{c}}
\cos^3{{\textstyle\frac{1}{2}}\bar{c}}-\sin^3{{\textstyle\frac{1}{2}}\bar{c}}
\cos{{\textstyle\frac{1}{2}}\bar{c}}\right)\otimes\varepsilon^3\\ 
\nonumber & & +O(\varepsilon^4)
\Big\} \hat{O}_A+ \\ \nonumber
 && \left\{\sin^2{{\textstyle\frac{1}{2}}\bar{c}}
\cos^2{{\textstyle\frac{1}{2}}\bar{c}}\otimes{\rm id}+ 
\left(\sin{{\textstyle\frac{1}{2}}\bar{c}}
\cos^3{{\textstyle\frac{1}{2}}\bar{c}}-
\sin^3{{\textstyle\frac{1}{2}}\bar{c}}\cos{{\textstyle\frac{1}{2}}\bar{c}}
\right)\otimes\varepsilon  \right. \\\nonumber
 && +{\textstyle\frac{1}{4}}\left(-6\sin^2{{\textstyle\frac{1}{2}}\bar{c}}
\cos^2{{\textstyle\frac{1}{2}}\bar{c}}+\cos^4{{\textstyle\frac{1}{2}}
\bar{c}}+\sin^4{{\textstyle\frac{1}{2}}\bar{c}}
\right)\otimes\varepsilon^2+{\textstyle\frac{2}{3}}\left(-\sin{{\textstyle\frac{1}{2}}\bar{c}}
\cos^3{{\textstyle\frac{1}{2}}\bar{c}}+\sin^3{{\textstyle\frac{1}{2}}\bar{c}}
\cos{{\textstyle\frac{1}{2}}\bar{c}}\right)\otimes\varepsilon^3\\ 
& & +O(\varepsilon^4) \Big\}
\hat{O}_c \bigg].
\end{eqnarray}

{}From (\ref{IsoHol}), the relevant combinations of sines and cosines on
the isotropic basis above have the matrix form
\begin{eqnarray}
\nonumber \langle\bar{\nu}|\sin^2{{\textstyle\frac{1}{2}}\bar{c}}
\cos^2{{\textstyle\frac{1}{2}}\bar{c}}|\bar{\mu}\rangle
 &=&
-2^{-4}\left(
\delta_{\n
,\bar{\mu}+4}-2\delta_{\n
,\bar{\mu}}+
\delta_{\n
,\bar{\mu}-4}\right), \\
\nonumber \langle\bar{\nu}|\sin{{\textstyle\frac{1}{2}}\bar{c}}
\cos^3{{\textstyle\frac{1}{2}}\bar{c}}|\bar{\mu}\rangle
&=&
-2^{-4}i\left(
\delta_{\n
,\bar{\mu}+4}+2\delta_{\n
,\bar{\mu}+2}-
2\delta_{\n
,\bar{\mu}-2}-\delta_{\n
,\bar{\mu}-4}\right),\\
\nonumber \langle\bar{\nu}|\sin^3{{\textstyle\frac{1}{2}}\bar{c}}
\cos{{\textstyle\frac{1}{2}}\bar{c}}|\bar{\mu}\rangle
&=&
2^{-4}i\left(
\delta_{\n
,\bar{\mu}+4}-2\delta_{\n
,\bar{\mu}+2}+
2\delta_{\n
,\bar{\mu}-2}-\delta_{\n
,\bar{\mu}-4}\right),\\
\nonumber \langle\bar{\nu}|\sin^4{{\textstyle\frac{1}{2}}\bar{c}}
|\bar{\mu}\rangle
&=& 2^{-4}\left(
\delta_{\n
,\bar{\mu}+4}-4\delta_{\n
,\bar{\mu}+2}+
6\delta_{\n
,\bar{\mu}}-4\delta_{\n
,\bar{\mu}-2}
+\delta_{\n
,\bar{\mu}-4}\right),\\
\langle\bar{\nu}|\cos^4{{\textstyle\frac{1}{2}}\bar{c}}|\bar{\mu}\rangle
&=& 2^{-4}\left(
\delta_{\n
,\bar{\mu}+4}+4\delta_{\n
,\bar{\mu}+2}+
6\delta_{\n
,\bar{\mu}}+4\delta_{\n
,\bar{\mu}-2}
+\delta_{\n
,\bar{\mu}-4}\right). \label{eq:sincosbarn}
\end{eqnarray}
So the total action in (\ref{eqH}) is given by
\begin{eqnarray}\label{eq:Hdelta}
\nonumber \langle\bar{\nu}|\hat{H}|\bar{\mu}\rangle&=& 
{\textstyle\frac{1}{4\pi}} i 
\gamma^{-3} G^{-1}
\ell_{\rm P}^{-2} \left\{\delta_{\n,\bar{\mu}}\left[ \left( 4-18\varepsilon^2
\right)\hat{O}_A(\bar{\nu})+2\hat{O}_c(\bar{\nu})  \right] \right. \\
\nonumber &+&  \delta_{\n,\bar{\mu}+4}\left[ \left( -2+2i\varepsilon
+\varepsilon^2 \right)\hat{O}_A(\bar{\nu})+ \left(-1-2i\varepsilon
+2\varepsilon^2 \right)
\hat{O}_c(\bar{\nu}) \right] \\
&+&  \left. \delta_{\n,\bar{\mu}-4}\left[ \left( -2-2i\varepsilon +\varepsilon^2
\right)\hat{O}_A(\bar{\nu})+ \left(-1+2i\varepsilon +2\varepsilon^2 \right)
\hat{O}_c(\bar{\nu}) \right]+O(\varepsilon^3) \right\} 
\end{eqnarray}
as matrix elements in ${\cal H}_{\rm iso}$ taking operator values in
${\cal H}_{\rm S}$ (for fixed $\bar{\nu}$, $\hat{O}_A(\bar{\nu})$ and
$\hat{O}_c(\bar{\nu})$ are considered as operators on
$|\bar{\nu}\rangle\otimes{\cal H}_{\rm S}\cong{\cal H}_{\rm S}$).

We can act with (\ref{eq:Hdelta}) on a generic state $|s>= \sum_{\nu}
s_{\nu}(\varepsilon) |\nu>$ and use
(\ref{eq:omna}, \ref{eq:omnc}). For this we write the Hamiltonian
in the form
 \begin{equation}\label{eq:evol}
 <\n ,\varepsilon | \hat{H}| s> =
-{\textstyle\frac{1}{16\pi}}\sqrt{\textstyle\frac{3}{2}} G^{-1} 
\gamma^{-3/2} \ell_{\rm P}\ 
\left\{ \hat{A}_{\n+4} s_{\n+4}(\varepsilon) + \hat{B}_{\n} s_{\n}
(\varepsilon) + \hat{C}_{\n-4} s_{\n-4}(\varepsilon) 
\right\},
\end{equation}
where
\begin{eqnarray} \label{n+4}
\nonumber \hat{C}_{\bar{\nu}}&=&
-{\textstyle\frac{2}{3}} (|\n|^{1/2}\sig_{3/2}(\n)+ |\n|
\Delta_3|\n|^{1/2})+ {\textstyle\frac{2}{3}}(|\n|^{1/2}\sig_{3/2}(\n)-
2|\n|\Delta_3|\n|^{1/2}) i\varepsilon\\
 && +  \left({\textstyle\frac{2}{3}} |\n|^{1/2} (\chi_{3/2}(\n)+\n^{-1}
\sig_{3/2}(\n))- \sig\n \Delta_3|\n|^{1/2}+
|\n|\Delta_3(|\n|^{-1/2}\sig\n) \right)\hat{P} \nonumber\\
 &&+ {\textstyle\frac{3}{2}} \left( -|\n|^{-1/2}\sig\n \chi_{3/2}(\n)+
{\textstyle\frac{1}{2}} |\n|^{-3/2}\sig_{3/2}(\n)+ \sig\n
\Delta_3(|\n|^{-1/2}\sig\n)+ {\textstyle\frac{1}{2}} |\n|
\Delta_3|\n|^{-3/2}\right) \hat{P}^2\nonumber\\
&&- \left(|\n|^{1/2}\chi_{3/2}(\n)+ |\n|^{-1/2}\sig\n\sig_{3/2}(\n)+
2\sig\n \Delta_3|\n|^{1/2}- 2|\n| \Delta_3(|\n|^{-1/2}\sig\n) \right)
i\varepsilon\hat{P} \nonumber\\
&&+ {\textstyle\frac{1}{3}} \left(|\n|^{1/2}\sig_{3/2}(\n)+
4|\n|\Delta_3|\n|^{1/2}\right) \varepsilon^2 +O(3)\,,
\end{eqnarray}
%
%
\begin{eqnarray}\label{n}
\nonumber \hat{B}_{\bar{\nu}}&=& {\textstyle\frac{4}{3}}
(|\n|^{1/2}\sig_{3/2}(\n)+ |\n|\Delta_3|\n|^{1/2})\\
 &&-2\left(|\n|^{1/2}\chi_{3/2}(\n)+ |\n|^{-1/2}\sig\n \sig_{3/2}(\n)-
\sig\n \Delta_3|\n|^{1/2}+ |\n|\Delta_3(|\n|^{-1/2}\sig\n) \right)
\hat{P} \nonumber\\
 &&+ 3\left(|\n|^{-1/2}\sig\n\chi_{3/2}(\n)- {\textstyle\frac{1}{2}}
|\n|^{-3/2} \sig_{3/2}(\n)- \sig\n \Delta_3(|\n|^{-1/2}\sig\n)-
{\textstyle\frac{1}{2}} |\n| \Delta_3|\n|^{-3/2} \right) \hat{P}^2
\nonumber\\
&& -6|\n|^{1/2}\sig_{3/2}(\n)\varepsilon^2 +O(3)
\end{eqnarray}
%
and $\hat{A}_{\bar{\nu}}=\overline{\hat{C}_{\bar{\nu}}}$. Note that
this is not the adjoint of $\hat{C}_{\bar{\nu}}$ but the complex
conjugate (acting only on the numerical coefficients, not on possible
factors of the imaginary unit in a derivative representation of
$\hat{P}$). Complex conjugation and adjoint are not the same in this
case because of the $\hat{\varepsilon}\hat{P}$-term. Also the whole
constraint operator is not self-adjoint since we started with a
non-symmetric ordering with holonomy operators to the left.

\section{Evolution}

The constraint equation given by the operator (\ref{eq:evol}), i.e.\
requiring physical states to be annihilated by the constraint
operator, results in a difference-differential equation for
$s_n(\varepsilon)$ which can be interpreted as evolution equation in the
volume $\bar{\nu}$ as internal time. This equation is
difference-differential because we keep the underlying discreteness
from quantum geometry in the isotropic variable, but treat the
anisotropy perturbatively. Also in isotropic models one can have
difference-differential equations if there is a matter degree of
freedom such as a scalar, but the crucial difference here is that
there are differential operators at all levels of the difference
equation and in particular the leading ones. This has implications for
the recurrence scheme since those operators have to be inverted in
some manner. We perform here only a basic analysis to check if the
scheme has satisfactory properties concerning evolution and relation
to classical behavior.

We focus on the stability issue for which in our context three
different notions occur. First, in the perturbative setting stability
is already relevant at the classical level where one needs
to make sure that anisotropies stay small enough for sufficiently long
time. As can be seen from the discussion leading to (\ref{pinternal}),
this can be ensured for some range of evolution at least in the vacuum
case for our situation. The second notion is stability at the quantum
level in the sense of \cite{FundamentalDisc}, requiring that there are
no exponentially growing solutions to the difference equation. This
will be the focus in the following sections. Finally, also numerical
stability issues can arise when difference or difference-differential
equations are solved numerically.

We consider the evolution equation (\ref{eq:evol}) without matter for
large $\n$, for which the constraint equation reduces to
\begin{eqnarray}\label{diff-recur}
\nonumber \frac{-\sqrt{3}\ \ell_{\rm P}}{2^{3/2}\ \kappa \ 
\gamma^{3/2}} \left[ \left( -\frac{4}{3}+\frac{4}{3}i \varepsilon +
\frac{2}{3} \varepsilon^2- \frac{6}{\n} i \varepsilon \hat{P} +\frac{3}{2\n^2 } 
\hat{P}^2\right) s_{\n +4}(\varepsilon) \right. &+&  \\
\nonumber \left( -\frac{4}{3}-\frac{4}{3}i \varepsilon +\frac{2}{3} 
\varepsilon^2+ \frac{6}{\n} i \varepsilon \hat{P} +\frac{3}{2\n^2} \hat{P}^2\right) 
s_{\n -4}(\varepsilon) &+&  \\
\left. \left( \frac{8}{3}-\frac{2}{\n} \hat{P} -12 \varepsilon^2-
\frac{3}{\n^2} 
\hat{P}^2 \right) s_{\n}(\varepsilon)\right]&=& 0
\end{eqnarray}

Now we propose a restricted function space for the solutions of the
form
\begin{equation} \label{restr}
 s_{\bar{\nu}}(\varepsilon)=p^{(k)}_{\bar{\nu}}(\varepsilon)\,
 e^{-(\varepsilon-\varepsilon_0)^2/4\sigma^2}, 
\end{equation}
where $p^{(k)}_{\bar{\nu}}$ are complex valued polynomials of order
$k$. Since we expanded the Hamiltonian constraint to second order, we
also should disregard terms of higher order in the solutions and work
with $k=2$. Then,
\begin{equation}\label{propoSol}
p_{\bar{\nu}}^{(2)} (\varepsilon)= A_0(\n)+A_1(\n) \ \varepsilon+
A_2(\n)\ \varepsilon^2
\end{equation}
where $A_0$, $A_1$ and $A_2$ are complex valued functions of the
discrete time $\n$ only, to be determined from initial values through
the difference equation. This class of functions is general enough for
our purposes because, even though $\varepsilon_0$ and $\sigma$ are fixed,
the Gaussian can be deformed and spread if $A_1$ and $A_2$ grow as
functions of $\n$. Moreover, to the given order of expansion also
non-zero expectation values for $\hat{p}_{\varepsilon}$ are allowed
through complex valued $A_1$ and $A_2$. For a Gaussian state peaked at
anisotropy $p_{\varepsilon}^0$ we need a phase factor $e^{6i\varepsilon
p_{\varepsilon}^0/\gamma\ell_{\rm P}^2}$ which for small enough
$p_{\varepsilon}^0$ can be expanded, only contributing imaginary parts to
$A_1$ and $A_2$. For this, we need to assume $p_{\varepsilon}^0\ll
\ell_{\rm P}^2$ which is possible but not guaranteed by our general
perturbation assumptions. Still, for the tests we are going to perform
such small values are sufficient. For more detailed aspects one has to
generalize the allowed class of functions.

Computing (\ref{diff-recur}) using
$\hat{P}=-\frac{2}{3}i\md/\md\varepsilon$ and (\ref{restr}),
(\ref{propoSol}) for large $\n$, and collecting terms of the same order
in $\varepsilon$ we obtain
\begin{eqnarray}\label{system1}
A_0(\n+4) \left( 2\sigma^2-8 \n^2 \sigma^4-\varepsilon_0^2 \right) -2 A_0(\n) \left( 2\sigma^2-8\n^2 \sigma^4-2i\n \sigma^2 \varepsilon_0-\varepsilon_0^2 \right) &&\\
\nonumber  +A_0(\n-4) \left( 2\sigma^2-8 \n^2 \sigma^4-\varepsilon_0^2 \right) &&\\
\nonumber  +A_1(\n+4) \left( -4 \sigma^2 \varepsilon_0 \right)-2 A_1(\n) \left( -4i\n \sigma^4-4 \sigma^2\varepsilon_0 \right) +A_1(\n-4) \left( -4 \sigma^2 \varepsilon_0 \right) &&\\
\nonumber  +A_2(\n+4) \left( -8\sigma^4 \right) -2 A_2(\n) \left( -8\sigma^4\right)
+ A_2(\n-4) \left( -8\sigma^4\right) &=&0
\end{eqnarray}
from leading order in $\varepsilon$,
\begin{eqnarray} \label{system1A1}
A_0(\n+4) \left( 8i\n^2 \sigma^4-12 \n \sigma^2 \varepsilon_0+2\varepsilon_0 \right) -2 A_0(\n) \left(-2i \n \sigma^2+2\varepsilon_0 \right) && \\
\nonumber   +A_0(\n-4) \left( -8i\n^2 \sigma^4+12 \n \sigma^2 \varepsilon_0+2\varepsilon_0 \right) &&\\
\nonumber  +A_1(\n+4) \left(6\sigma^2-24 \n \sigma^4-8\n^2 \sigma^4- \varepsilon_0^2 \right) -2 A_1(\n) \left(6\sigma^2-8\n^2\sigma^4-\varepsilon_0^2-2i\n \sigma^2 \varepsilon_0 \right) && \\
\nonumber   +A_1(\n-4) \left(6\sigma^2+24 \n \sigma^4-8\n^2 \sigma^4- \varepsilon_0^2 \right) &&\\
\nonumber  +A_2(\n+4) \left( -8\sigma^2 \varepsilon_0 \right) -2 A_2(\n) \left( -8i\n \sigma^4-8 \sigma^2 \varepsilon_0 \right)
+ A_2(\n-4) \left( -8\sigma^2 \varepsilon_0 \right) &=&0
\end{eqnarray}
from linear order, and
\begin{eqnarray} \label{system1A2}
 A_0(\n+4) \left( -1+12\n \sigma^2+4\n^2 \sigma^4 \right) -2 A_0(\n)
\left(-1+36\n^2 \sigma^4 \right) 
&&\\\nonumber   
+A_0(\n-4) \left( -1-12\n \sigma^2+4\n^2 \sigma^4 \right)&&\\
\nonumber  +A_1(\n+4) \left(8i\n^2 \sigma^4+2\varepsilon_0-12\n \sigma^2\varepsilon_0 \right) -2 A_1(\n) \left(2i\n \sigma^2+2\varepsilon_0 \right) &&\\
\nonumber   +A_1(\n-4) \left(-8i\n^2 \sigma^4+2\varepsilon_0+12\n \sigma^2\varepsilon_0 \right) &&\\
\nonumber  +A_2(\n+4) \left( 10\sigma^2-48\n \sigma^4-8\n^2 \sigma^4- \varepsilon_0^2 \right) -2 A_2(\n) \left(10\sigma^2-8\n^2 \sigma^4-2i\n \sigma^2 \varepsilon_0-\varepsilon_0^2 \right)&&\\ \nonumber
  +A_2(\n-4) \left( 10\sigma^2+48\n \sigma^4-8\n^2 \sigma^4- \varepsilon_0^2 \right)&=&0
\end{eqnarray}
from quadratic order.
As the equations are quite
complicated, we discuss here mainly their approximation for large
volume which can already be used to see relations to the classical
behavior.

\subsection{Asymptotic $\n$ solution}

While the perturbation scheme does not require a relation between the
magnitudes of $\bar{\nu}$ and $\varepsilon_0$ or $\sigma\sim\varepsilon_0\ll
1$, for a given choice of $\varepsilon_0$ and $\sigma$ consistent with
the perturbation assumptions we can look at the asymptotic form of the
equations where only highest powers of $\bar{\nu}$ are used. For the
second order equations, this requires $\bar{\nu}\gg \sigma^{-2}$ which
is our assumption for the asymptotic regime. Simplifying
the equations then leads to
\begin{eqnarray}\label{system2}
\nonumber A_0(\n+4)-2 A_0(\n)+A_0(\n-4)&=& 0, \\
\nonumber A_1(\n+4)-2 A_1(\n)+A_1(\n-4)&=& i 
\left[ A_0(\n+4)-A_0(\n-4) \right],\\
\nonumber A_2(\n+4)-2 A_2 (\n)+A_2(\n-4)&=&\frac{1}{2} 
\left[ A_0(\n+4)-18 A_0(\n)+A_0(\n-4) \right] +\\
 & &i \left[ A_1(\n+4)-A_1(\n-4) \right].
\end{eqnarray}
We can already observe that, in contrast to (\ref{system1}), these
equations are independent of the value of $\varepsilon_0$ in complete
analogy to our previous observation for the asymptotic classical
solution (\ref{pinternal}). Also, the equations have imaginary terms
which is necessary for $p_{\varepsilon}$ to change in internal time
(in the $\varepsilon$-representation chosen here, $p_{\varepsilon}$
enters through the phase of the wave function). Even if
$p_{\varepsilon}^0$ is initially zero, i.e.\ $A_0$, $A_1$ and $A_2$
are all real, the wave function must become complex after some steps
such that anisotropy also in the triad arises automatically.

Before solving (\ref{system2}) we generalize it to allow dust like
matter (matter Hamiltonian independent of $\n$) such that the middle
coefficients on the left hand sides of the equations are $2\rho$ with
a constant $\rho\leq 1$, reproducing the vaccum case above for
$\rho=1$.  We then put the system in a first order form, considering
solutions only on the lattice $4{\mathbb Z}$ which is sufficient to
see the behavior of long-time evolution. Accordingly, we introduce the
vector
\begin{displaymath}\label{vector}
\mathbf{v}(\n+4)=\left( 
\begin{array}{c} A_0(\n+4)\\  A_1(\n+4)\\ A_2(\n+4)\\A_0(\n)\\ 
A_1(\n)\\ A_2(\n) 
\end{array} \right).
\end{displaymath}
With this definition we can cast (\ref{system2}) in a
first-order linear matrix equation (of stepsize four):
\begin{equation} \label{LinearSyst}
M\ \mathbf{v}(\n+4)+ N\ \mathbf{v}(\n)= \mathbf{0}\qquad \mbox{ or } \qquad
\mathbf{v}(\n+4)+ Q\ \mathbf{v}(\n)= \mathbf{0},
\end{equation}
where the matrix $Q$ is 
\begin{displaymath}
Q= \left( \begin{array}{cccccc}
-2\rho & 0& 0& 1& 0& 0 \\
-2i \rho & -2\rho & 0 & 2i & 1& 0\\
9+\rho & 2i\rho & -2\rho& -2& 2i & 1\\
-1 & 0 & 0 & 0 & 0 & 0\\
0& -1 &0 & 0 & 0 & 0\\
0 & 0 & -1 & 0 & 0 & 0
\end{array} \right).
\end{displaymath}  

It is possible to bring this matrix to its Jordan canonical form (see
App.~\ref{AppJordan}) $J= S^{-1} Q S$ where

\begin{displaymath}
J= \left( \begin{array}{cccccc}
\lambda_1 & 1& 0&0 & 0& 0 \\
0 &  \lambda_1& 1 & 0 & 0& 0\\
0 & 0&\lambda_1 & 0& 0 & 0\\
0 & 0 & 0 &\lambda_2 & 1 & 0\\
0& 0 &0 & 0 & \lambda_2 & 1\\
0 & 0 & 0 & 0 & 0 & \lambda_2
\end{array} \right)\,.
\end{displaymath}  
Here, $\lambda_{1,2}= \pm \sqrt{\rho^2-1}-\rho$ are the two
eigenvalues of $Q$ with multiplicity three (which degenerate further
in the vaccum case $\rho=1$).

{}From here we can obtain the solution starting from
initial conditions in (\ref{LinearSyst}):
\begin{equation}
\mathbf{v}(\n_0+4k)= Q^k \mathbf{v}(\n_0).
\end{equation}
We obtain
\begin{equation} \label{solution}
Q^k = S J^k S^{-1},
\end{equation}
where $J$ is the Jordan form for $Q$ above and each one of its
block powers is
\begin{equation} \label{powersJ}
J_i^k = \sum_{j=0}^{2} \binom{k}{j} \lambda_i^{k-j} \ N^j.
\end{equation}
$N$ is the off-diagonal term in $J$. 

We obtain the transition matrix $S$ by computing the generalized
eigenvectors associated to $\lambda_{1,2}$, which we do not write
down; however we write the general $k$-power of $J$ above:
\begin{displaymath}
J^k= \left( \begin{array}{cccccc}
\lambda_1^k & k \lambda_1^{k-1} & \binom{k}{2}\lambda_1^{k-2} & 0 & 0 & 0\\
0 & \lambda_1^k & k \lambda_1^{k-1} & 0 & 0 & 0 \\
0 & 0 & \lambda_1^k & 0 & 0 & 0\\
0 & 0 & 0 & \lambda_2^k & k\lambda_2^{k-1} & \binom{k}{2}\lambda_2^{k-2} \\
0 & 0 & 0 & 0 & \lambda_2^k & k\lambda_2^{k-1} \\
0 & 0 & 0 & 0 & 0 & \lambda_2^k \end{array} \right).
\end{displaymath}

{}From this the solution for the $A$ functions is
\begin{eqnarray} \label{soln>>1}
A_0(\n_0+4k) &=& \frac{1}{2}(\rho^2-1)^{-1/2}
\left((-\lambda_1^k\lambda_2+  \lambda_1\lambda_2^k)
A_0(\n_0)
 + (\lambda_2^k -\lambda_1^k) A_0(\n_0-4)\right)\,,\\
\nonumber A_1(\n_0+4k) &=& \frac{1}{2}(\rho^2-1)^{-1/2}
\left((-\lambda_1^k\lambda_2+\lambda_1\lambda_2^k) A_1(\n_0)
- ik( -\lambda_1^2\lambda_2^{k-1} +\lambda_1^{k-1}\lambda_2^2)
A_0(\n_0)\right. \\
&&+ \left.i (\lambda_2^k - 
\lambda_1^k -k\lambda_1^{k-1}\lambda_2
+k\lambda_2^{k-1} \lambda_1) A_0(\n_0-4)
 (\lambda_2^k -\lambda_1^k) A_1(\n_0-4)\right)
\end{eqnarray}
and
\begin{eqnarray}
%
\nonumber A_2(\n_0+4k) &=& \frac{1}{4}\left[ \frac{9}{(\rho^2-1)^{3/2}} \left( \lambda_1^k-\lambda_2^k \right) \right. \\
\nonumber &+& k\lambda_1^{k-1} \left( \frac{-9-2\rho+18 \rho^2+2 \rho^3}{\rho^2-1}+\frac{-1+18\rho+2\rho^2}{\sqrt{\rho^2-1}} \right)  \\
\nonumber &+& \binom{k}{2} \frac{2\lambda_1^{k-2}}{\rho^2-1} \left( 1-5\rho^2+4\rho^4-3\rho \sqrt{\rho^2-1}+4 \rho^3 \sqrt{\rho^2-1} \right)  \\
\nonumber &+& k\lambda_2^{k-1}\left(\frac{ -9-2\rho+18\rho^2+2\rho^3}{\rho^2-1}+\frac{-1+18 \rho+3\rho^2+18 \rho^3-2 \rho^4}{(\rho^2-1)^{3/2}} \right) \\
\nonumber &+& \left. \binom{k}{2} \lambda_2^{k-2} \left( \frac{2-12\rho^2+18\rho^4-8\rho^{6}}{(\rho^2-1)^2} +\frac{6\rho -14\rho^3 +8 \rho^{5}}{(\rho^2-1)^{3/2}} \right) \right] A_0(\n_0) \\
\nonumber &-& \frac{ik}{2} \left( \frac{ \lambda_1^{k-1} \lambda_2^2-\lambda_2^{k-1} \lambda_1} {\sqrt{\rho^2-1}} \right) A_1(\n_0)
+ \frac{1}{2}\left( \frac{-\lambda_1^k \lambda_2+ \lambda_2^k \lambda_1 }{\sqrt{\rho^2-1}} \right) A_2(\n_0) \\
\nonumber &+& \frac{1}{4} \left[ \lambda_1^k \frac{ 1-9\rho+\rho^2 }{(\rho^2-1)^{3/2}}+\lambda_2^k \frac{ 1+9\rho-\rho^2 }{(\rho^2-1)^{3/2}}  \right. \\
\nonumber &+& 3k\left( -\lambda_1^{k-1} \left\{1+\frac{3\rho}{\rho^2-1} +\frac{3+\rho}{\sqrt{\rho^2-1}} \right\} +\lambda_2^{k-1} \left\{ -1-\frac{3\rho}{\rho^2-1}+\frac{3+\rho}{\sqrt{\rho^2-1}} \right\} \right) \\
\nonumber &+& \binom{k}{2} \left( \lambda_1^{k-2} \left\{ 4\rho+2\frac{2\rho^2-1}{\sqrt{\rho^2-1}} \right\} \right. \\
\nonumber &+&\left. \left. \lambda_2^{k-2} \left\{ \frac{4\rho}{(\rho+1)^2}-2\frac{2\rho^2-1}{(\rho+1)^{3/2} \sqrt{\rho-1}}\right\} \right) \right] A_0(\n_0-4) \\
\nonumber &+& \frac{i}{2}\left( \frac{\lambda_2^k-\lambda_1^k}{(\rho^2-1)^{\frac{3}{2}}}
- k \frac{\lambda_1^{k-1}\lambda_2-
\lambda_1\lambda_2^{k-1}}{\sqrt{\rho^2-1}}\right)
A_1(\n_0-4)\\
&+& \frac{\lambda_2^k - \lambda_1^k}{2\sqrt{\rho^2-1}} A_2(\n_0-4).
\end{eqnarray}
This solution is valid only for $|\rho| \neq 1$, for otherwise
the eigenvalues would be completely degenerate.

So we have obtained the solution to the evolution equation in the
asymptotic $\n$ regime. We should note that the solutions grow as
powers of the order $k$ (no larger than second) as a consequence of
degeneracy, although the leading coefficient $A_0$ is bounded.

\subsection{Inclusion of matter}

One can hope that matter terms remove the degeneracy present for the
case for pure gravity, which similarly happens in the isotropic model
(see \cite{Sing} for an explicit isotropic vacuum solution which is
also unbounded). As the previous discussion showsm however, this is
not the case for dust which has a constant diagonal Hamiltonian, $\hat{H}=
E \mathbb{I}$.

As the next simple possibility we can consider a cosmological constant
term
\begin{equation}\label{matterHamilt}
\hat{H}_{\textrm{matter}} = \Lambda \hat{V}.
\end{equation}
Using (\ref{eq:volmn}) the evolution equation for large $\n$
(\ref{diff-recur}) modifies to
\begin{eqnarray}\label{evolLargeN}
\nonumber   \left( -\frac{4}{3}+\frac{4}{3}i \varepsilon +\frac{2}{3} \varepsilon^2- \frac{6}{\n} i \varepsilon \hat{P} +\frac{3}{2\n^2 } \hat{P}^2\right) s_{\n +4}  &+&  \\
\nonumber \left( -\frac{4}{3}-\frac{4}{3}i \varepsilon +\frac{2}{3} \varepsilon^2+ \frac{6}{\n} i \varepsilon \hat{P} +\frac{3}{2\n^2} \hat{P}^2\right) s_{\n -4} &+&  \\
\nonumber  \left( \frac{8}{3}-\frac{2}{\n} \hat{P} -12 \varepsilon^2- \frac{3}{\n^2} \hat{P}^2 \right) s_{\n} &=& \\
-\frac{4}{3}\pi G \gamma^3\Lambda  \left( \frac{2 \n^{3/2}}{3} -\frac{9}{4 \sqrt{\n}}\hat{P}^2 \right) s_{\n} & & . 
\end{eqnarray}
With the same ansatz (\ref{restr}), (\ref{propoSol}) the right
hand side of each equation becomes
\begin{equation}\label{system1Hamilt}
 \frac{\n^{3/2}}{6}\left[ A_0(\n) (54\sigma^2+32\n^2
\sigma^4-27 \varepsilon_0^2) + A_1(\n)(-108\sigma^2 \varepsilon_0)+
A_2(\n)(-216 \sigma^4) \right]
\end{equation}
for (\ref{system1}),
\[
 \frac{\n^{3/2}}{6}\left[ A_0(\n)(54\varepsilon_0)
+A_1(\n)(162\sigma^2+32\n^2 \sigma^4-27 \varepsilon_0^2)+
A_2(\n)(-216\sigma^2 \varepsilon_0) \right]
\]
for (\ref{system1A1}) and
\[
  \frac{\n^{3/2}}{6}\left[
A_0(\n)(-27)+A_1(\n)(54\varepsilon_0)+A_2(\n) (270\sigma^2+32\n^2
\sigma^4-27 \varepsilon_0^2) \right]
\]
for (\ref{system1A2}).

It is easier to check the degeneracy in the asymptotic $\n$ regime: in
this case the right hand sides of (\ref{system2}) acquire an extra
term $\frac{2^7}{9}\pi G \gamma^3 \Lambda \n^{3/2} A_i(\n)$,
respectively. Coefficients in the asymptotic difference equation are
thus no longer constant and we do not find explicit solutions. Still,
around any fixed $\bar{\nu}$ eigenvalues remain degenerate.

\subsection{Large $\n$ behavior}

Rather than looking at more involved matter choices, we now use only
large $\bar{\nu}$ and keep next order terms in $\bar{\nu}^{-1}$. This
gives rise to more complicated correction terms in (\ref{system1})
compared to the asymptotic equations, changing all
coefficients. Rather than trying to solve this second order system of
difference equations, we perform a local stability analysis around
large fixed $\bar{\nu}$ similar to that in the previous section

To this end we put the system (\ref{diff-recur}) again in a first
order form (\ref{LinearSyst}), where now the matrices $\mathbf{M,N,Q}$
are functions of $(\n, \varepsilon_0, \sigma)$. In this case the
stability of the system is obtained if and only if the module of the
eigenvalues of the matrix $\mathbf{Q}$ above are less or equal to
one. However its characteristic polynomial is of order six and,
recalling the form of (\ref{system1}), each one of its coefficients is
a very complex function of $(\n, \varepsilon_0, \sigma)$. Therefore we
perform a numerical analysis to see the behavior of each one of the
coefficient functions $A_i(\n)$ in (\ref{propoSol}).

The characteristic polynomial is of the form
\begin{eqnarray}\label{characterPol}
\nonumber \lambda^6+ h_5 \lambda^5+ h_4 \ \lambda^4+ h_3 \ \lambda^3+ h_2 \ \lambda^2+ h_1 \ \lambda+ h_0&=&0, \\
h_j = h_j(\n, \varepsilon_0, \sigma), \quad j=0 \ldots 5 &&
\end{eqnarray}
which one can solve numerically to find choices of $\varepsilon_0$ and
$\sigma$ where all six roots have an absolute value less than or equal to
one.



\section{Perturbations and the singularity}
\label{s:Sing}

When perturbing in basic variables, the expansion is valid only for
$p_{\varepsilon}\ll\bar{p}$ and thus breaks down very close to the
classical singularity. Indeed, the main aim for studying perturbations
is to understand effects in structure formation where one needs at
least perturbative inhomogeneities for which anisotropies serve as a
model. This is usually modelled in semiclassical regimes, in an
inflationary phase or sometimes through a bounce at relatively large
volume, far away from the Planck scale.  It is nevertheless
instructive to discuss implications for the singularity issue itself
close to vanishing $\bar{p}$, as they might appear in a perturbative
setting.

In the asymptotic $\bar{\nu}$ equation (\ref{system2})
we saw that the relevant matrix $M$ in (\ref{LinearSyst}) at
highest order of the difference equation is always invertible such
that evolution cannot break down in this regime. But
this is not guaranteed at small $\bar{\nu}$ where a breakdown of
evolution can occur even at $\bar{\nu}>0$ where one would not expect
singularities classically. There are even divergences in coefficients
of the difference equation such as in $\hat{C}_{\bar{\nu}}$ for
$\bar{\nu}=3$ coming, e.g., from $\Delta_3|\bar{\nu}|^{-3/2}$. This is
certainly a small value for $\bar{\nu}$, but note that $p_{\varepsilon}$
can be small compared to $\bar{p}$ even for $\bar{\nu}$ of the order
one which perturbatively is consistent using the fact that the
perturbative operator $p_{\varepsilon}$ has a continuous spectrum. As
discussed earlier, small $p_{\varepsilon}$ do not automatically bring us
into discrete regimes, unlike small $\bar{p}$. One can thus set up the
perturbative evolution such that all assumptions for the perturbation
scheme are satisfied, and yet the evolution breaks down at non-zero
volume. (Starting at large volume, the perturbative approximation is
expected to break down much before the classical singularity is
reached. But generically, a breakdown of evolution cannot be ruled out
since there are initial conditions for which the evolution breaks down
{\em before} the perturbative scheme ceases to make sense.) 

There is thus no removal of singularities in the perturbative
model. Even worse, the perturbative equation could make us believe
that the non-singular evolution of the isotropic model is very special
and could not extend to less symmetric models, owing to the fact that
for $\varepsilon=P=0$ all divergent coefficients disappear. In such a
situation, it is necessary to use the full model from which the
perturbative equation is derived in order to see whether or not it
breaks down. As it turns out, for loop quantum cosmology the evolution
is non-singular even for anisotropic models, in contrast to the
perturbative appearance.

A different source for a breakdown of evolution appears because
coefficients of the difference equation are now differential operators
on the perturbative degree of freedom. These operators must be
inverted to proceed with the evolution, which is not generically
possible and may require special boundary conditions. Note that, if a
physical inner product is not known, even generic invertibility would
not be enough: one needs to make sure that any initial state can be
evolved without breakdown to conclude singularity freedom. Even if
almost all initial values would lead to non-singular evolution, all of
them could be ruled out by the physical inner product (see also
\cite{DegFull}).  We do not discuss this in detail here but only
conclude that, while the fully quantized anisotropic model is
non-singular, there is no such statement in the perturbative
quantization. 

It is interesting to compare this situation with recent results in
string theory \cite{StringInHom} where inhomogeneities on a background
are seen to prevent the occurence of a bounce instead of a singularity
(which would otherwise be possible in the corresponding homogeneous
model). This looks similar to our perturbative quantization of
anisotropies on an isotropic background, even though the models and
techniques are certainly very different. In contrast, non-perturbative
background independent models studied so far are non-singular,
including inhomogeneous ones which classically have local physical
degrees of freedom
\cite{SphSymmSing}.

There is a further disadvantage of using perturbative or semiclassical
degrees of freedom in order to discuss the singularity issue, as for
instance suggested in \cite{BoundCoh}. Such a perturbative treatment
is unlikely to remain valid close to a singularity where potentially
all degrees of freedom can be excited strongly. Perturbations
certainly allow one to include all degrees of freedom which is
important for phenomenology, but properties of the singularity can be
extremely blurred as we have seen here. In contrast, symmetric models
completely remove many degrees of freedom which may have to be
reinserted later on when discussing their effect on evolution. But
symmetric models themselves are often classically singular, and those
singularities can be studied by quantizations of models. With the
models now available, the most characteristic types of classical
singularities can be studied also at the quantum level. This can then
show how classical singularities can be resolved by quantization, and
there is indeed a general mechanism in loop quantum gravity.

While the fundamental singularity issue concerning the extension of
solutions to the Hamiltonian constraint equation cannot be analyzed
perturbatively, this is still possible for phenomenological models of
bounces as studied in \cite{Bounces}
and used for the construction of oscillatory universe models in
\cite{Scen}.
Such bounces need to happen sufficiently far above the Planck scale in
order to ensure the validity of effective equations
\cite{Eff},
and thus avoid a potential breakdown of the perturbative discrete
evolution. This will also be true for perturbative analyses of
inhomogeneities. Also at the inhomogeneous level, one can employ
approximate but non-perturbative methods in the spirit of the BKL
picture \cite{BKL}. In contrast to perturbative schemes, this can then
also be applied to the fundamental singularity issue, as done in the
loop quantum gravity context with preliminary investigations in
\cite{NonChaos}.

\section{Conclusions}

We have mainly illustrated the reduction procedure to symmetric
situations available for loop quantum gravity, and contrasted it with
a new method for perturbative quantum degrees of freedom in the case
of anisotropy. We consider this mainly as a model situation for the
cosmologically more interesting but technically much more complicated
case of perturbative inhomogeneities. This already allowed us to draw
some cautionary conclusions for the singularity problem when discussed
in a perturbative setting, as well as for the relation between
symmetric models and perturbative less symmetric ones.

The resulting evolution equations are of a new type compared to other
models in loop quantum gravity. Although not studied in detail here,
they are treatable at least by numerical schemes and provide an
interesting setting to understand evolution equations in general and
numerical techniques to solve them. New schemes for an efficient
solution still need to be developed, also taking into account the
issue of pre-classicality \cite{DynIn} ignored here. This condition
requires solutions not to vary strongly on the scale of discreteness.
It can be expected for semiclassical behavior, but is also necessary
to ensure the correct behavior of wave packets. Techniques to extract
pre-classical solutions, such as those based on generating functions
\cite{GenFunc}, continued fractions \cite{ContFrac} or numerical ones
using suitable function bases \cite{PreClassBasis}, have been
introduced but are not yet available for the equations considered
here.

Also the equations themselves can be generalized in several
directions. Within the same degree of symmetry one can discuss
different matter choices and, at the level of the ansatz
(\ref{restr}), more general classes for the functional behavior of the
wave function depending on anisotropy. Moreover, different orderings
of the constraint operator, most importantly the symmetric one, can be
tried. The bigger step consists in including inhomogeneous degrees of
freedom perturbatively for which the analysis in this paper serves as
preparation. Classically, the split into symmetric and non-symmetric
variables would be given by a mode decomposition with respect to the
symmetric background. A priori, there are no insurmountable
difficulties in repeating the analysis here for this situation, even
though technically it would be much more involved. Inhomogeneities can
already be included explicitly by using midi-superspace models for
which the loop framework has been constructed in
\cite{SphSymm,SphSymmHam}. These situations already allow the
inclusion of inhomogeneities of cosmological interest.

Our analysis also shows that the reduced constraint of the more
symmetric model still plays a role for the less symmetric one and
is only amended by perturbative correction terms. Those terms are derived by
expanding the less symmetric constraint operator, and not by
quantizing the classically expanded constraint in the fashion of
\cite{Halliwell}. Indeed, the additional terms are much more contrived
than expected classically. Although more complicated, these terms are
derived from the less-symmetric model and are thus
well-motivated. After a detailed study of the evolution, such as
restrictions from quantum stability, one can draw conclusions for the
form of the less symmetric constraint operator. (Such restrictions
indeed arise easily as the closed isotropic model, \cite{IsoCosmo}
compared to \cite{Closed}, and the more detailed analysis in
\cite{FundamentalDisc} showed.) Eventually, with a better
understanding of the relation between models and the full theory also
at the level of the Hamiltonian constraint, one can then hope to
restrict possible full constraint operators and reduce their
ambiguities.

\section*{Acknowledgements}

We thank Jonathan Engle for discussions prior to publication of
\cite{SymmQFT}. MB is grateful to the Isaac Newton Institute for
Mathematical Sciences, Cambridge for hospitality during the workshop
``Global Problems in Mathematical Relativity,'' where this paper was
completed, and thanks the organizers Piotr Chrusciel and Helmut
Friedrich for an invitation. HH was supported by the fellowship
A/04/21572 of Deutscher Akademischer Austauschdienst (DAAD). HAMT and
HH acknowledge partial support from Mexico's National Council of
Science and Technology (CONACyT), under grants CONACyT-40745-F,
CONACyT-NSF-E-120-0837.

\section*{Appendix}

\appendix

\section{Approximate action of operators on peaked states}
\label{ApproxOp}

For the perturbative expansion of the Hamiltonian constraint at the
quantum level it is necessary to define and discuss an appropriate
notion of expansions of operators. We are not interested in expansions
valid for any state of the Hilbert space but only in expansions that
can be used for states corresponding to small anisotropy such as those
used in (\ref{restr}). Thus, the expansion parameter is not contained
directly in the original operator but introduced through a restricted
class of states considered for the action. We then start by choosing a
space ${\cal D}_{\delta}$ of states in the original Hilbert space
${\cal H}$, or even a subset of a cylindrical space ${\rm Cyl}_{\rm
S}$, depending on the intended perturbation parameter $\delta$ (or
several ones such as $\varepsilon_0$ and $p_{\varepsilon}^0$ in the main
part of this paper, and define
\begin{defi}
  An operator $\hat{O}_k$ on ${\cal H}$ is an {\em approximation of
  order $k$} to an operator $\hat{O}$ if $\hat{O}_k\Psi$ agrees, with
  respect to the inner product, with $\hat{O}\Psi$ for all
  $\Psi\in{\cal D}_{\delta}$ up to order $\delta^{k+1}$, i.e.\
\[
 \langle\Psi'|(\hat{O}-\hat{O}_k)\Psi\rangle=O(\delta^{k+1})
\]
for all $\Psi'\in{\rm Cyl}_{\rm S}$. Agreement with respect to the inner
product in this sense will be denoted by $\hat{O}_k\Psi \sim
\hat{O}\Psi$.
\end{defi}

To show that this is realized for the expansions performed for
anisotropies in this paper, we choose ${\cal D}_{q_0,p^0}$ to be the
span of states of the form $\Psi(q)=\sigma^{-1/2}s(q)
\exp(-(q-q_0)^2/4\sigma^2) \exp(ip^0q)$ for a given
$\sigma<q_0$ and a polynomial $s$ (we do not normalize states but keep
track of factors of $\sigma$). Expansions performed here always lead
to approximations $\hat{O}_k$ which are polynomials in the basic
operators $\hat{q}$ and $\hat{p}$ of the
anisotropy. This implies that any $\hat{O}_k$ considered here maps the
space ${\cal D}_{\delta}$ to itself and can be used to show that the
validity of expansions is preserved by taking products:
\begin{lemma}
 Let $\hat{O}^{(1)}_k$ and $\hat{O}^{(2)}_k$ be approximations of
$\hat{O}^{(1)}$ and $\hat{O}^{(2)}$, respectively, such that they fix
the spaces ${\cal D}_{\delta}$ and ${\rm Cyl}_{\rm S}$. Then
$\hat{O}^{(1)}_k\hat{O}^{(2)}_k$ is an approximation of
$\hat{O}^{(1)}\hat{O}^{(2)}$.
\end{lemma}
\begin{proof}
By assumption we have
$\hat{O}^{(1)}\hat{O}^{(2)}\Psi\sim \hat{O}^{(1)}\hat{O}^{(2)}_k\Psi$
because we can simply replace $\Psi'$ by $\hat{O}^{(1)\dagger}\Psi'$ in
the definition of agreement. Since $\hat{O}^{(2)}_k\Psi$ is in ${\cal
D}_{\delta}$ for $\Psi\in{\cal D}_{\delta}$, we have
\[
 \hat{O}^{(1)}\hat{O}^{(2)}\Psi\sim
 \hat{O}^{(1)}_k\hat{O}^{(2)}_k\Psi
\]
which finishes the proof.
\end{proof}

We can therefore split the expansion into different steps as done in
the main calculations here. This also simplifies the proof that the
performed expansion of the Hamiltonian constraint is an approximation
because we only need to show that multiplication operators can be
approximated by simply expanding the multiplying function, which we
would apply for our calculations in both the coordinate and momentum
representation. We then only need
\begin{lemma}
 Let $\hat{O}\Psi(q)=O(q)\Psi(q)$ be a multiplication operator on
 $\hat{H}=L^2({\mathbb R},\md q)$. Any partial Taylor sum
 $O_k(q)=\sum_{n=0}^k \frac{1}{n!} O^{(n)}(0) q^n$ gives rise to an
 approximation $\hat{O}_k\Psi(q)=O_k(q)\Psi(q)$ of $\hat{O}$ order $k$
with respect to ${\cal D}_{\delta}$ as above.
\end{lemma}
\begin{proof}
 Let $\Psi(q)=\sigma^{-1/2}s(q)\exp(-(q-q_0)^2/4\sigma^2)$ and $\Psi'\in{\rm
 Cyl}_{\rm S}$. We have
\begin{eqnarray*}
 |\langle\Psi'| (\hat{O}-\hat{O}_k)\Psi\rangle| &=& 
\sigma^{-1/2}\left|\int\md q
\bar{\Psi}'(q) (O(q)-O_k(q))s(q)\exp(-(q-q_0)^2/4\sigma^2)\right|\\
 & \leq&
2C_1\sigma^{-1/2}\int_{q_0}^{q_0+1} \exp(-(q-q_0)^2/4\sigma^2) q^{k+1} \md q\\
&&+
2C_2\sigma^{-1/2}\int_{q_0+1}^{\infty} \exp(-(q-q_0)^2/4\sigma^2) q^{k+1} \md q
\end{eqnarray*}
with positive constants $C_1$ and $C_2$ depending only on the supremum
of the real and imaginary parts of $\bar{\Psi}'(q)s(q)$. The first
integral can be estimated by
\begin{eqnarray*}
 \int_{q_0}^{q_0+1} \exp(-(q-q_0)^2/4\sigma^2) q^{k+1} \md q &=&
\int_0^1 \exp(-q^2/4\sigma^2) (q+q_0)^{k+1} \md q\\
&=&
\sigma^{k+2}\int_0^{\sigma^{-1}} \exp(-q^2/4)(q+q_0/\sigma)^{k+1}\md q\\
& \leq&
\sigma^{k+2}\int_0^{\infty} \exp(-q^2/4)(q+q_0/\sigma)^{k+1}\md q
\end{eqnarray*}
where the last integral exists and does not depend strongly on $q_0$
and $\sigma$ since they are assumed to be of the same order. The whole
expression is thus of order $\sigma^{k+2}$. For the second integral,
we have
\begin{eqnarray*}
 \int_{q_0+1}^{\infty} \exp(-(q-q_0)^2/4\sigma^2) q^{k+1} \md q &=&
e^{-(2\sigma)^{-2}} \int_0^{\infty} \exp(-(q^2+2q)/4\sigma^2)
(q+q_0+1)^{k+1} \md q\\
& \leq& e^{-(2\sigma)^{-2}} \int_0^{\infty} 
\exp(-q^2/4\sigma^2) (q+2)^{k+1} \md q
\end{eqnarray*}
where the last integral is a polynomial in $\sigma$ and the whole
expression is thus of the order $e^{-(2\sigma)^{-2}}$ times a
polynomial in $\sigma$. By our assumptions on the states in ${\cal
D}_{\sigma}$, we have $\sigma$ of the order or smaller than $q_0$
which concludes the proof.
\end{proof}

\section{Jordan canonical form}
\label{AppJordan}

\subsection{Generalized eigenvectors and transition matrix}

In this section we briefly review the Jordan canonical form of a
square complex matrix. For further details and the proofs of the
theorems see for example \cite{alglin}. We begin stating the following

\begin{theo}
If a square matrix $A$ of order $n$ has $s$
linearly independent eigenvectors, then it is similar to a matrix $J$ of
the following form, called a {\em Jordan canonical form}
\begin{equation} \label{jordan1}
J= S^{-1} A S = \left( \begin{array}{cccc}
J_1 & & & 0\\
 & J_2 & & \\
 &  & \ddots & \\
0 & & & J_s \end{array} \right),
\end{equation}
in which each $J_i$, called the {\em Jordan block}, is a triangular
matrix of the form \begin{equation} \label{jordan2} J_i= \left(
\begin{array}{cccc}
\lambda_i &1 & & 0\\
 & \ddots &\ddots & \\
 &  & \ddots & 1\\
0 & & & \lambda_i \end{array} \right),
\end{equation}
where $\lambda_i$ is a single eigenvalue of $A$. Each Jordan block
corresponds to a linearly independent eigenvector.
\end{theo}

This theorem states that we can bring any square matrix to its Jordan
form. The usefulness of this form lies in the fact that it is easier
to obtain powers of the matrix $A$, as can be seen by taking a look
at the form of $J$ in (\ref{jordan1}), (\ref{jordan2}). The matrix $S$ in
(\ref{jordan1}) is called the {\em transition matrix}. Now we state
the procedure to obtain both matrices $J$ and $S$.

The basic elements to obtain the Jordan form are the {\em generalized
eigenvectors}, replacing ordinary eigenvectors of a diagonalizable
matrix.

\begin{defi}
A nonzero vector $\mathbf{x}$ is said to be a {\em generalized
eigenvector} of $A$ of rank $k$ belonging to an eigenvalue $\lambda$
if
\begin{equation}
\left( A-\lambda I \right)^k \mathbf{x} = \mathbf{0} \quad \textrm{and}  \quad \left( A-\lambda I \right)^{k-1} \mathbf{x} \neq \mathbf{0}.
\end{equation}
\end{defi}
If $k=1$ this is the usual definition of an eigenvector. For a
generalized eigenvector $\mathbf{x}$ of rank larger than one belonging
to the eigenvalue $\lambda$ define
\begin{eqnarray}
\nonumber \mathbf{x}_k &=& \mathbf{x}, \\
\nonumber \mathbf{x}_{k-1} &=& (A-\lambda I)\mathbf{x} \ \ =\ (A-\lambda I)\mathbf{x}_k , \\
\nonumber \mathbf{x}_{k-2} &=& (A-\lambda I)^2 \mathbf{x} \ =\ (A-\lambda I)\mathbf{x}_{k-1} , \\
\nonumber \vdots &  \\
\nonumber \mathbf{x}_{2} &=& (A-\lambda I)^{k-2} \mathbf{x} \ =\ (A-\lambda I)\mathbf{x}_{3} ,\\
 \mathbf{x}_{1} &=& (A-\lambda I)^{k-1} \mathbf{x} \ =\ (A-\lambda I)\mathbf{x}_{2}.
\end{eqnarray}
One can check that each one of the $\mathbf{x}_j$ above is a generalized eigenvector of the eigenvalue $\lambda$. 
\begin{defi}
The set of vectors $\{ \mathbf{x}_1, \ldots , \mathbf{x}_k \}$ is
called a {\em chain of generalized eigenvectors} belonging to the
eigenvalue $\lambda$.
\end{defi}

Now the transition matrix $S$ can be constructed from the chains of
linearly independent generalized eigenvectors of $A$ \cite{alglin}.

\subsection{Powers of square matrices}

The Jordan canonical form of any square matrix $A$ enables us to
compute its powers $A^k$. Let $J$ be the Jordan canonical form of an
arbitrary $n\times n$ matrix $A$ such that
\begin{displaymath}
S^{-1}AS= J = \left( \begin{array}{ccc}
J_1 & & \\
 & \ddots & \\
 &  & J_s \end{array}
\right),
\end{displaymath}
where $S$ is made of the generalized eigenvectors of $A$ and $J_s$
are Jordan blocks.

Since we have 
 \begin{displaymath}
A^k=S^{-1}J^k S = S \left( \begin{array}{ccc}
J_1^k & & \\
 & \ddots & \\
 &  & J_s^k \end{array}
\right) S^{-1}
\end{displaymath}
for $k=1,2,\ldots$, it is enough to compute $J^k$ for each Jordan
block $J$. An $m\times m$ Jordan block $J$ belonging to an eigenvalue
$\lambda$ of $A$ may be written as
\begin{displaymath}
J= \left( \begin{array}{cccc}
\lambda & 1 &  & 0 \\
0 & \ddots & \ddots & \\
 &  & \lambda & 1 \\
0 &  & 0 & \lambda \end{array} \right) =
\lambda \mathbb{I}_{m\times m} + \left( 
\begin{array}{cccc}
0 & 1 & \cdots & 0 \\
0 & \ddots & \ddots & 0 \\
\vdots &  & 0 & 1 \\
0 & \cdots & 0 & 0 \end{array}
\right) = \lambda \mathbb{I}+N,
\end{displaymath} 
where $\mathbb{I}$ is the identity matrix and $N$ is
nilpotent. Clearly we have
\begin{equation}
J^k = \left( \lambda \mathbb{I}+N \right)^k = \sum_{j=0}^k \binom{k}{j} \lambda^{k-j} N^j.
\end{equation}

But $N^k=0$ for $k \geq m$ and thus, by defining $\binom{k}{l}=0$ if $k<l$
\begin{eqnarray}
\nonumber J^k &=& \sum_{j=0}^{m-1} \binom{k}{j} \lambda^{k-j} N^j \\
 & =& \lambda^k \mathbb{I} + k \lambda^{k-1}N + \cdots + \binom{k}{m-1} 
\lambda^{k-(m-1)} N^{m-1}.
\end{eqnarray}


\end{document}